\documentclass[a4paper,11pt]{article}
\usepackage{natbib}
\usepackage[colorlinks,citecolor=blue,urlcolor=blue]{hyperref}
\usepackage{epsf}
\usepackage{epsfig}
\usepackage{multirow}
\usepackage[english]{babel}
\usepackage{graphicx}
\usepackage{geometry}
\usepackage{natbib}
\usepackage{amssymb}
\usepackage{amsmath}
\usepackage{mathtools}
\mathtoolsset{showonlyrefs=true}
\usepackage{color}
\usepackage{enumerate}
\usepackage{authblk}
\usepackage[table,xcdraw,dvipsnames]{xcolor}
\providecommand{\keywords}[1]{\textbf{\textit{Keywords---}} #1}
\usepackage{booktabs}
\usepackage{makecell}

\usepackage[font={small,it}]{caption}
\usepackage[labelformat=simple]{subcaption}
\usepackage{bm}

\newcommand{\specificthanks}[1]{\@fnsymbol{#1}}

\title{Variational Inference for \\ the Latent Shrinkage Position Model}
\author[]{Xian Yao Gwee\thanks{xian-yao.gwee@ucdconnect.ie} }
\author[]{Isobel Claire Gormley\thanks{claire.gormley@ucd.ie}}
\author[]{Michael Fop\thanks{michael.fop@ucd.ie\\[0.2cm] Isobel Claire Gormley and Michael Fop have contributed equally to this work.}}
\affil[]{School of Mathematics and Statistics,\protect \\ University College Dublin.}

\date{\today}

\begin{document}

\maketitle

\begin{abstract}
The latent position model (LPM) is a popular method used in network data analysis where nodes are assumed to be positioned in a $p$-dimensional latent space. The latent shrinkage position model (LSPM) is an extension of the LPM which automatically determines the number of effective dimensions of the latent space via a Bayesian nonparametric shrinkage prior. However, the LSPM reliance on Markov chain Monte Carlo for inference, while rigorous, is computationally expensive, making it challenging to scale to networks with large numbers of nodes. We introduce a variational inference approach for the LSPM, aiming to reduce computational demands while retaining the model's ability to intrinsically determine the number of effective latent dimensions. 
The performance of the variational LSPM is illustrated through simulation studies and its application to real-world network data. To promote wider adoption and ease of implementation, we also provide open-source code.

\end{abstract}


\section{Introduction}
\label{sec:vlspm-intro}

In recent years network data have become increasingly prevalent and complicated, appearing for example as neural connections in neuroscience \citep{aliverti2019spatial}, as trade relationships \citep{lyu_2023_latent} and as the spread of diseases in epidemiology \citep{chu_2021_dynamic}. Such networks can have hundreds to thousands of nodes, making them challenging to model using current statistical tools. Thus, there is a need for statistical network models that can be used for inference on large networks in a computationally efficient manner.

The latent position model \citep[LPM,][]{hoff_2002_latent} is a popular network model due to its ease of interpretation. The LPM assumes that nodes have positions in a $p$-dimensional latent space, and that the distances between the nodes' latent positions influence the observed edge formation process. The latent shrinkage position model (LSPM) \citep{gwee_2022_a} extended the LPM to allow for automatic inference on the number of effective dimensions of the latent space.
This is achieved through a Bayesian nonparametric framework where a multiplicative truncated gamma process (MTGP) prior is assumed on the variance of the latent positions. The MTGP prior induces shrinkage on the variance of the latent positions as the number of dimensions increases. This approach eliminates the requirement to fit multiple LPMs and then choose between them using model selection criteria to identify the optimal $p$.

In the Bayesian setting, inference for the LPM and its extensions has typically relied on Markov chain Monte Carlo (MCMC) techniques, for example in the clustering setting \citep{handcock_2007_modelbased}, in longitudinal network models \citep{chu_2021_dynamic}, and in multiview network data \citep{dangelo_2023_modelbased}, among others. These sampling-based methods, while rigorous, have a high computational cost. As the number of nodes increases, their computational cost increases quadratically 
\citep{saltertownshend_2013_variational, rastelli_2018_computationally}, making the LPMs challenging to scale to large networks. Since the LSPM uses Metropolis-within-Gibbs sampling, it also suffers from this issue.



Various efforts have been made to improve computational efficiency when inferring LPMs via MCMC sampling methods, including the adoption of Hamiltonian Monte Carlo \citep{spencer_2022_faster}, noisy MCMC \citep{rastelli_2018_computationally}, and the use of case-control approximate likelihood \citep{raftery_2012_fast, sewell_2015_latent}, each having varying degrees of success. Instead of employing such sampling methods, here a variational inference (VI) approach  \citep{jordan_1999_an, blei_2017_variational} is adopted in which the posterior distribution is approximated via an optimization process, reducing the computational cost of inference. The idea stems from literature where variational inference has been successfully employed for computationally efficient estimation of LPMs. For example, \cite{saltertownshend_2013_variational} introduced a variational Bayes algorithm for the latent position cluster distance model, while \cite{gollini_2016_joint} introduced a variational inference algorithm for a latent position distance model for multilayer networks. In the context of longitudinal networks, both  \cite{sewell_2017_latent} and \cite{liu_2023_variational} used variational inference, with a focus on the latent position projection model and the distance model, respectively. Additionally, \cite{aliverti_2021_stratified} presented a stratified stochastic variational inference method for high-dimensional network factor models. Many variational algorithms have exhibited empirical success, and recent theoretical contributions by \cite{yang_2020_alpha} and \cite{liu_2023_variational} have provided deeper insights into the statistical guarantees of these methods.

Here, we propose the {\em variational inference LSPM} (VI-LSPM), a variational inference approach for the LSPM to perform posterior inference for large scale networks. While \cite{saltertownshend_2013_variational} and \cite{gollini_2016_joint} have developed a variational inference approach for the LPM under standard parametric priors, here we propose a variational inference approach for the LPM where a Bayesian nonparametric prior, specifically the MTGP prior, is employed. Our simulation studies show improved computational cost compared to MCMC approaches, while the ability to identify the number of effective dimensions is retained. We also apply VI-LSPM to analyse a cat brain connectivity binary network with a moderate number of nodes and moderate density, and a relatively large but sparse binary network describing the nervous system of a worm. 

In what follows, Section \ref{sec:vlspm-model} provides an introduction to the LSPM and outlines a VI algorithm for posterior inference. Sections \ref{sec:vlspm-sim} and \ref{sec:vlspm-appl} illustrate the performance of the proposed method with simulation studies and real-world examples, respectively. Section \ref{sec:vlspm-disc} concludes with a discussion. To aid practitioners in employing the VI-LSPM, the necessary R code is available in the \href{https://gitlab.com/gwee95/lspm}{\texttt{lspm}}  GitLab repository.

\section{Variational inference for the latent shrinkage position model}
\label{sec:vlspm-model}
\subsection{The latent shrinkage position model} \label{ssec:vlspm-lspm}

Consider a binary network with $n$ nodes, encoded in a $n \times n$ adjacency matrix, $\mathbf{Y}$, where entry $y_{i,j} = 1$ if there is an edge between nodes $i$ and $j$, and $0$ otherwise, for $i, j = \{1, \ldots, n\}$. Self-edges are not considered. 

In the LPM \citep{hoff_2002_latent} each node has an unobserved latent position $\mathbf{z}_i$, in a $p$-dimensional Euclidean latent space. Edges are assumed independent, conditional on the latent positions of the nodes, giving the sampling distribution:
$$
    \mathbb{P}(\mathbf{Y}\mid\alpha, \mathbf{Z}) = \prod_{i \neq j} \mathbb{P}(y_{i,j}\mid\alpha, \mathbf{z}_i,\mathbf{z}_j)
$$
where $\mathbf{Z}$ is the
matrix of latent positions and $\alpha$ is a global parameter that captures the overall connectivity level in the network. A logistic model formulation is then employed to characterize the edge-formation process, where the log odds of an edge between nodes $i$ and $j$ depends on the squared Euclidean distance \citep{gollini_2016_joint,gwee_2022_a} between their latent space positions $\mathbf{z}_i$ and $\mathbf{z}_j$ i.e., 
\begin{equation} \label{eq:vlspm-logitprob}
\log \frac{q_{i,j}}{1-q_{i,j}} = \alpha - \Vert \mathbf{z}_i-\mathbf{z}_j \Vert^2_2  \end{equation} 
where $q_{i,j}$ represents the probability of an edge between nodes $i$ and $j$.

The LSPM \citep{gwee_2022_a} extends the LPM through a Bayesian nonparametric approach which enables automatic inference of $p$, the number of effective latent space dimensions necessary to fully describe the network. The latent positions are assumed to lie in an infinite dimensional space and to have a Gaussian distribution with zero mean and a diagonal precision matrix $\mathbf{\Omega}$, whose entries $\omega_{\ell}$ denote the precision of the latent positions in dimension $\ell$, for $\ell=1, \ldots, \infty$. A shrinkage prior, the multiplicative truncated gamma process (MTGP), is assumed for the precision parameters, ensuring shrinkage of the variance in higher dimensions. Each latent dimension, $h = 1, \ldots, \infty$, has an associated shrinkage strength parameter, $\delta_h$ and the precision $\omega_{\ell}$ in dimension $\ell$ is the cumulative product of $\delta_1$ to $\delta_{\ell}$, i.e., for $i = 1, \ldots, n$,
\begin{equation}
\begin{aligned}  
\label{eq:vlspm-z_mvn}
    \mathbf{z}_{i} \sim \text{MVN}(\mathbf{0}, \mathbf{\Omega}^{-1}) \qquad \mathbf{\Omega} = \begin{bmatrix} 
    \omega_{1}^{-1} & \dots & 0 \\
    \vdots & \ddots & \\
    0 &        & \omega_{\infty}^{-1}
    \end{bmatrix}  \qquad
    \omega_{\ell} = \prod_{h=1}^{\ell} \delta_{h} \quad \mbox{ for } \ell = 1, \ldots, \infty.  \\
\end{aligned}
\end{equation}

Gamma priors, $ \text{Gam}(a_1, b_1)$, are assumed for the shrinkage strength parameters, where for dimensions $h>1$, a truncated gamma prior, $ \text{Gam}^{\text{T}}(a_2, b_2, t_2) $, with a left truncation point at $t_2 = 1$ is assumed to ensure shrinkage, leading to the MTGP prior, i.e.
\begin{equation}
\begin{aligned}  
\notag
    \delta_1 \sim \text{Gam}(a_1, b_1=1) \qquad
    \delta_{h} \sim \text{Gam}^{\text{T}}(a_2, b_2=1, t_2 = 1) \quad \mbox{ for }h  > 1. \\
\end{aligned}
\end{equation}
Under this MTGP prior, the variances of the latent positions in the higher dimensions will shrink towards zero causing the dimensions with non-negligible variance to correspond to the effective dimensions necessary to describe the network.

The joint posterior distribution of the LSPM is
\begin{equation}
\begin{aligned} 
\label{eq:vlspm-post}
\mathbb{P}(\alpha, \mathbf{Z}, \bm{\delta} \mid \mathbf{Y})  \propto \mathbb{P}(\mathbf{Y}\mid\alpha, \mathbf{Z}) \mathbb{P}(\alpha) \mathbb{P}(\mathbf{Z} \mid \bm{\delta}) \mathbb{P}(\bm{\delta}). 
\end{aligned}
\end{equation}
In \cite{gwee_2022_a}, inference on the LSPM relies on MCMC which, while rigorous, is computationally expensive and makes fitting the LSPM to networks with large numbers of nodes challenging in practice. Here we propose a variational approach to inference for the LSPM, providing a fast but approximate solution, making application of the LSPM to large networks computationally feasible. 

\subsection{Variational inference for the LSPM}
\label{sec:variational}
Variational inference (VI) is a computational technique used to approximate complex, intractable posterior distributions with simpler, tractable ones (see \cite{blei_2017_variational} for a comprehensive review). Instead of directly computing the posterior, VI reformulates the problem as an optimization task. This is done by selecting a family of simpler distributions to optimize the parameters based on the Kullback-Leibler (KL) divergence, such that the constructed variational distribution is as close as possible to the true posterior. A widely-adopted technique for VI is the mean-field approach \citep{consonni_2007_meanfield}, which represents the posterior distribution using a completely factorized form. This allows for the simplification of the expectations needed for the calculation of the KL divergence. 

Although the MTGP prior is nonparametric, it is not implementable in practice without setting a finite truncation level $p$ on the number of dimensions fitted. Hence, for fixed $p$, we apply the VI approach to the LSPM posterior \eqref{eq:vlspm-post} by defining the variational posterior as
$$
\mathbb{Q}(\alpha, \mathbf{Z}, \bm{\delta})  = \mathbb{Q}(\alpha) \prod_{i=1}^n \mathbb{Q}(\mathbf{z}_i)\mathbb{Q}(\delta_1) \prod_{h=2}^p \mathbb{Q}(\delta_h),
$$
where $\mathbb{Q}(\alpha) = N(\tilde{\mu}_{\alpha},\tilde{\sigma}^2_{\alpha}), \mathbb{Q}(\mathbf{z}_i) = \text{MVN} (\tilde{\mathbf{z}}_i, \tilde{\bm{\Omega}}^{-1}), \mathbb{Q}(\delta_1) = \text{Gam} (\tilde{a}_1,\tilde{b}_1)$, and  $\mathbb{Q}(\delta_h)=\text{Gam}^{T}(\tilde{a}_2^h,\tilde{b}_2^h, 1)$. The variational parameters are distinguished with a tilde for clarity and are collectively denoted by $\tilde{\Theta} = \{\tilde{\mu}_{\alpha},\tilde{\sigma}^2_{\alpha}, \tilde{\mathbf{z}}_i, \tilde{\bm{\Omega}}^{-1}, \tilde{a}_1,\tilde{b}_1, \tilde{a}_2^h,\tilde{b}_2^h \}
$. The variational parameter $\tilde{\mathbf{\Omega}}^{{-1}}$ has diagonal entries taking the inverse of the cumulative products of the expected values of $\delta_1$ and $\delta_{\ell}$ under the variational distributions $\mathbb{Q}(\delta_1)$ to $\mathbb{Q}(\delta_{\ell})$.

Typically, the KL divergence is used as a measure of closeness to minimise the distance between the variational posterior $\mathbb{Q}(\alpha, \mathbf{Z}, \bm{\delta})$ and the true posterior $\mathbb{P}(\alpha, \mathbf{Z}, \bm{\delta}  \mid \mathbf{Y})$:
$$
\begin{aligned}
\mathrm{KL}[\mathbb{Q}(\alpha, \mathbf{Z}, \bm{\delta}) \| \mathbb{P}(\alpha, \mathbf{Z}, \bm{\delta} \mid \mathbf{Y})] & =-\int \mathbb{Q}(\alpha, \mathbf{Z}, \bm{\delta}) \log \frac{\mathbb{P}(\alpha, \mathbf{Z}, \bm{\delta} \mid \mathbf{Y})}{\mathbb{Q}(\alpha, \mathbf{Z}, \bm{\delta})} d(\alpha, \mathbf{Z}, \bm{\delta}) \\
&=-\int \mathbb{Q}(\alpha, \mathbf{Z}, \bm{\delta}) \log \frac{\mathbb{P}(\mathbf{Y}, \alpha, \mathbf{Z}, \bm{\delta})}{\mathbb{Q}(\alpha, \mathbf{Z}, \bm{\delta})} d(\alpha, \mathbf{Z}, \bm{\delta}) \\
& \qquad +\log \mathbb{P}(\mathbf{Y}).
\end{aligned}
$$
Note that minimising the KL divergence is equivalent to maximising the evidence lower bound (ELBO) with respect to the variational distribution \citep{bishop_2006_pattern}. 

The KL divergence between the variational posterior and the true posterior for the LSPM can be written as:
\begin{equation}
\begin{aligned} \label{eq:fullKL}
\mathrm{KL}&[\mathbb{Q}(\alpha, \mathbf{Z}, \bm{\delta}) \| \mathbb{P}(\alpha, \mathbf{Z}, \bm{\delta} \mid \mathbf{Y})] \\
= &  \operatorname{KL}\left[\mathbb{Q}(\alpha) \| \mathbb{P}(\alpha)\right]+\sum_{i=1}^{n} \operatorname{KL}\left[\mathbb{Q}\left(\mathbf{z}_{i}\right) \| \mathbb{P}\left(\mathbf{z}_{i}\right)\right] \\
& +\operatorname{KL}\left[\mathbb{Q}(\delta_1) \| \mathbb{P}(\delta_1)\right]+\sum_{h=2}^p \operatorname{KL}\left[\mathbb{Q}\left(\delta_{h}\right) \| \mathbb{P}\left(\delta_{h}\right)\right]\\
& -\mathbb{E}_{\mathbb{Q}(\mathbf{Z}, \alpha \mid \mathbf{Y})}\left\{\log \left[\mathbb{P}(\mathbf{Y} \mid \alpha, \mathbf{Z}, \bm{\delta})\right]\right\}  +\log \mathbb{P}(\mathbf{Y}) \\
= &-\frac{1}{2} \log \left( \frac{\tilde{\sigma}_\alpha^2}{\sigma_\alpha^2 }\right)-\frac{1}{2} +\frac{\tilde{\sigma}_\alpha^2}{2\sigma_\alpha^2} +\frac{(\tilde{\mu}_\alpha-\mu_\alpha)^2}{2\sigma_\alpha^2}   \\
& -\frac{np}{2} + \frac{n}{2} \log \operatorname{det}(\mathbf{\Omega}^{-1} \tilde{\mathbf{\Omega}})   +\frac{1}{2} \sum_{i=1}^{n} \tilde{\mathbf{z}}_i^{\top} \mathbf{\Omega} \tilde{\mathbf{z}}_i +\frac{n}{2} \operatorname{tr}(\mathbf{\Omega} \tilde{\mathbf{\Omega}}^{-1}) \\
&+ \tilde{a}_1  \left[\psi(\tilde{a}_1)+\frac{b_1}{\tilde{b}_1}+1\right] - a_1  \left[\psi(\tilde{a}_1)+\log(\frac{b_1}{\tilde{b}_1})\right] - \log\frac{\Gamma(\tilde{a}_1)}{\Gamma(a_1)} \\
& +\sum_{h=2}^{p} \operatorname{KL}\left[\mathbb{Q}\left(\delta_{h}\right) \| \mathbb{P}\left(\delta_{h}\right)\right]  -\mathbb{E}_{\mathbb{Q}(\mathbf{Z}, \alpha \mid \mathbf{Y})}\left\{\log \left[\mathbb{P}(\mathbf{Y} \mid \mathbf{Z}, \alpha)\right]\right\}  +\log \mathbb{P}(\mathbf{Y}) , \\
\end{aligned}
\end{equation}
where $\Gamma$ and $\psi$ are the gamma and digamma functions respectively (see Appendix \ref{app:vlspm-deriveKL} for full derivations). 
As in \cite{gollini_2016_joint}, the expected log-likelihood $\mathbb{E}_{\mathbb{Q}(\mathbf{Z}, \alpha \mid \mathbf{Y})}\{\log [\mathbb{P}(\mathbf{Y} \mid \mathbf{Z}, \alpha)]\}$ is approximated using the Jensen's inequality:

$$
\begin{aligned}
\mathbb{E}_{\mathbb{Q}(\mathbf{Z}, \alpha \mid \mathbf{Y})}\{\log [\mathbb{P}(\mathbf{Y} \mid \mathbf{Z}, \alpha)]\} &=\sum_{i \neq j}^{n} y_{i,j} \mathbb{E}_{\mathbb{Q}(\mathbf{Z}, \alpha \mid \mathbf{Y})}\left[\alpha-\left\|\mathbf{z}_{i}-\mathbf{z}_{j}\right\|^{2}\right] \\
& \quad -\mathbb{E}_{\mathbb{Q}(\mathbf{Z}, \alpha \mid \mathbf{Y})}\left[\log \left(1+\exp \left(\alpha-\left\|\mathbf{z}_{i}-\mathbf{z}_{j}\right\|^{2}\right)\right)\right] \\
& \leq \sum_{i \neq j}^{n} y_{i,j}\left(\mathbb{E}_{\mathbb{Q}(\mathbf{Z}, \alpha \mid \mathbf{Y})}\left[\alpha-\left\|\mathbf{z}_{i}-\mathbf{z}_{j}\right\|^{2}\right]\right) \\
& \quad -\log \left(1+\mathbb{E}_{\mathbb{Q}(\mathbf{Z}, \alpha \mid \mathbf{Y})}\left[\exp \left(\alpha-\left\|\mathbf{z}_{i}-\mathbf{z}_{j}\right\|^{2}\right)\right]\right) \\
& =\sum_{i \neq j}^{n} y_{i,j}\left(\tilde{\mu}_{\alpha}-2 \operatorname{tr}(\tilde{\mathbf{\Omega}}^{-1})-\left\|\tilde{\mathbf{z}}_{i}-\tilde{\mathbf{z}}_{j}\right\|^{2}\right) \\
& \quad -\log \left\{1+\frac{\exp \left(\tilde{\mu}_{\alpha}+\frac{1}{2} \tilde{\sigma}_{\alpha}^{2}\right)}{\operatorname{det}(\mathbf{I}+4 \tilde{\mathbf{\Omega}}^{-1})^{\frac{1}{2}}} \right. \\
& \qquad \left. \exp \left[-\left(\tilde{\mathbf{z}}_{i}-\tilde{\mathbf{z}}_{j}\right)^{\top}(\mathbf{I}+4 \tilde{\mathbf{\Omega}}^{-1})^{-1}\left(\tilde{\mathbf{z}}_{i}-\tilde{\mathbf{z}}_{j}\right)\right]\right\} .
\end{aligned}
$$
In contrast to \cite{saltertownshend_2013_variational}, which employs three first-order Taylor expansions to approximate the log-likelihood, the squared Euclidean distance formulated in \eqref{eq:vlspm-logitprob} benefits from using just one application of Jensen's inequality, yielding a comparatively tighter bound. \cite{jaakkola_2000_bayesian} offers an alternative approach to approximate the expected log-likelihood but, as it requires further approximations, it is more difficult to compute.

\subsubsection{Updating the variational parameters}
\label{ssec:vlspm-update}
The variational parameters $ \{ \tilde{\mu}_{\alpha},\tilde{\sigma}^2_{\alpha}, \tilde{\mathbf{z}}_i \}$ are updated by minimising \eqref{eq:fullKL} via a bracketing-and-bisection method and a conjugate gradient routine (see Appendix \ref{app:vlspm-partial} for a summary and Appendix \ref{app:vlspm-fullopt} for the full derivations).
However, the variational parameters $\{\tilde{a}_1,\tilde{b}_1, \tilde{a}_2^h,\tilde{b}_2^h \}$ for $h = 2, \ldots, p$ are updated analytically by identifying their optimal expressions as the gamma (for $h=1$) and truncated gamma (for $h>1$) densities respectively, with their own shape and rate parameters (see Appendix \ref{app:vlspm-partial} for a summary and Appendix \ref{app:vlspm-fullopt} for the full derivations). The variational inference algorithm then updates each of the variational parameters in turn, while holding the others constant.

While \cite{gollini_2016_joint} employed both first and second order Taylor expansions to obtain an analytical form for updating the variational parameters, poor performance was observed empirically when the dimension of the latent space was higher than two. Instead, we adopt the numerical optimization methods of \cite{saltertownshend_2013_variational} where the latent position variational means $\tilde{\mathbf{z}}_i$ are optimized via a conjugate gradient routine \citep{fletcher_1964_function} while the $\tilde{\mu}_{\alpha}$ and $\tilde{\sigma}^2_{\alpha}$ hyperparameters are optimized via a bracketing-and-bisection algorithm \citep{burden_1985_numerical}.

To initialize the estimation algorithm, here, we follow the strategy for obtaining starting values as taken in  \cite{gwee_2022_a}, with the exception that the truncation level is set as $p = 5$, which was found to work well in practice for the networks examined here. A well-chosen initial set of variational parameters is essential for the VI approach due to the potential presence of multiple local minima in the Kullback–Leibler divergence. To address the issue of local minima, a range of starting values is used. To do so, random noise is added to the initial latent positions obtained from multidimensional scaling, for a pre-specified value of $p$, distributed as $\text{MVN}(0, r^2)$ where $r^2$ is the five hundredths of the empirical variance of the multidimensional scaling positions. Ten random initialisations are considered and the best estimate is selected as the one with the highest ELBO value. 

Convergence of the VI algorithm can be assessed in a number of different ways. It is typically achieved  by monitoring the difference between the ELBO at successive iterations and stopping the algorithm when this difference is smaller than a pre-specified threshold. In our implementation, we monitor the approximate expected log-likelihood across iterations as it often makes the largest contribution to the evaluation of the ELBO, simplifying the calculations for the convergence check. In the settings considered here, convergence was typically achieved after a low number ($<50$) of iterations, using a threshold of 0.01.  

\section{Simulation studies}
\label{sec:vlspm-sim}
The performance of the VI-LSPM is assessed on simulated data scenarios with a computer equipped with an i7-10510U CPU and 16GB RAM. Performance of the VI-LSPM is assessed in 5 ways: computing time, Procrustes correlation (PC), the area under the receiver operating characteristic curve (AUROC), the area under the precision recall curve (AUPR)\, and through posterior predictive checks. Time is measured in seconds with shorter time indicating preferred performance. Procrustes correlation \citep{mardia_2008_multivariate} measures the similarity between different latent space configurations by minimising the sum of squared differences after Procrustes rotations, with higher values (up to a maximum of 1) indicating better correspondence. Here the PC is calculated between the estimated latent positions and the true positions, conditioning on the smaller number of dimensions between the two configurations compared. The AUROC and AUPR measure the goodness-of-fit of the model by comparing the estimated edge probabilities (calculated using the estimated model parameters and latent positions as in \eqref{eq:vlspm-logitprob}) with the observed data, with value of 1 implying perfect model fit. An AUROC value of 0.5 implies random predictions, while the AUPR's baseline value depends on the density of the network. In networks with low density, the AUPR will generally be lower than AUROC as it is based on the proportion of present edges.
Comparisons are made between the VI-LSPM and the MCMC-based LSPM (MCMC-LSPM), both implemented in the R package available in the \href{https://gitlab.com/gwee95/lspm}{\texttt{lspm}} repository.

Posterior predictive checks are conducted (in both the simulation studies and in the real networks considered in Section \ref{sec:vlspm-appl}) through simulating posterior predictive networks under the fitted VI-LSPM and comparing them with the relevant observed network. Similar to \cite{gwee_2022_a}, similarity metrics (accuracy and $F_1$ score), distance between networks (Hamming distance), and network properties (density and transitivity) are considered in the posterior predictive checks where an accuracy or $F_1$ score of one implies perfect fit, as does a Hamming distance of zero. Similarity between the posterior predictive network statistics of density and transitivity and their observed values also indicates good model fit. Posterior predictive checking of the VI-LSPM is performed with the estimate from the final iteration, while for the MCMC-LSPM it is performed via sampling the posterior distribution 30 times.



The simulated data are generated by Bernoulli trials for each node pair based on the probabilities derived from the distances between the nodes' latent positions. The latent positions are simulated according to \eqref{eq:vlspm-z_mvn} with the shrinkage strengths being manually set to explore their effect across different settings. Hyperparameters are set as $\mu_{\alpha} = 0, \sigma_{\alpha}=3, a_1=2, a_2=3, b_1 = b_2 = 1$. A total of 30 networks are simulated in each case. The VI-LSPM is run from 10 different initial positions on each of the 30 networks, with the run with the highest ELBO used for inference and performance assessment for each network.

The simulation studies are structured as follows: Section \ref{ssec:vlspm-truncation} examines the performance and impact of different truncation levels $p$ and Section \ref{ssec:vlspm-size} assesses the performance of the VI-LSPM under different network sizes $n$. 


\subsection{Study 1: truncation level}
\label{ssec:vlspm-truncation}

Here the effect of different truncation levels of the latent dimension is explored. Networks with $n=100$ are generated with the true number of latent dimensions $p^*=4$ and shrinkage strengths of $\delta_1=0.5$, $\delta_2=1.1$, $\delta_3=1.05$, and $\delta_4=1.15$. 
A value $\alpha=6$ is used, leading to networks with moderate density of $\simeq 20\%$. Three truncation levels are considered: $p = \{2, 4, 10\}$, representing situations where the truncation level has been underestimated, correctly specified, and overestimated, respectively. 

Table \ref{tab:sim_trunc_time} shows that the VI-LSPM converges up to two orders of magnitude faster than the MCMC-LSPM. Following \cite{gwee_2022_a}, 1,000,000 iterations were run for the MCMC approach. Although the time per iteration is higher for VI-LSPM due to more complex computations, the substantial reduction in the number of iterations results in lower computational cost.

\begin{table}[htb]
\centering
\caption{Comparison of VI-LSPM and MCMC-LSPM average model run time in seconds for different truncation levels $p$. Standard deviations are given in brackets.}
\label{tab:sim_trunc_time}
\begin{tabular}{|c|cc|}
\hline
                      & \multicolumn{2}{c|}{Time (sec)}                                   \\ \cline{2-3} 
\multirow{-2}{*}{$p$} & \multicolumn{1}{c|}{VI}                               & MCMC      \\ \hline
2                     & \multicolumn{1}{c|}{{ 23 (2)}}    & 2188 (57) \\ \hline
4                     & \multicolumn{1}{c|}{{ 54 (22)}}   & 2381 (38) \\ \hline
10                    & \multicolumn{1}{c|}{{ 142 (139)}} & 2471 (89) \\ \hline
\end{tabular}
\end{table}

Table \ref{tab:sim_trunc_compare} demonstrates good performance of the VI-LSPM when compared with the truth, and and also when compared with the MCMC-LSPM solutions. Table \ref{tab:sim_trunc_compare} shows good correspondence between estimated and true configurations, with PC values between the true latent positions and the VI-LSPM estimated latent positions of at least 0.68. Comparison between VI-LSPM estimated latent positions and the MCMC-LSPM estimated latent positions also shows good correspondence, with PC values of at least 0.827. For both $p=4$ and $p=10$, the Procrustes correlation is similar when VI-LSPM is compared to both the truth and the MCMC-LSPM estimates, indicating that a higher number of dimensions does not highly affect the estimated latent positions configuration. The AUROC shows good model fit with values of more than 0.82 (when compared to the truth) and 0.81 (when compared to the MCMC approach) across the different truncation levels but with $p = \{4, 10\}$ showing better and similar fit. Similarly, the AUPR shows good model fit for $p = \{4, 10\}$ as it takes values much higher than the network density. These results suggest the use of $p > p^*$ is preferable to $p < p^*$.  

\begin{table}[htb]
\centering
\caption{Performance of the VI-LSPM for different truncation levels $p$, assessed through Procrustes correlation (PC), the area under the receiver operating characteristic curve (AUROC), and the area under the precision recall curve (AUPR) through comparison with the truth and with the MCMC-LSPM estimates. Standard deviations are given in brackets.}
\label{tab:sim_trunc_compare}
\resizebox{\linewidth}{!}{
\begin{tabular}{|c|ccc|lll|}
\hline
                      & \multicolumn{3}{c|}{Comparison with truth}                                                                                                                 & \multicolumn{3}{c|}{{ Comparison with MCMC}}                                                                                                       \\ \cline{2-7} 
\multirow{-2}{*}{$p$} & \multicolumn{1}{c|}{PC}                                 & \multicolumn{1}{c|}{AUROC}                                & AUPR                                 & \multicolumn{1}{c|}{{ PC}}          & \multicolumn{1}{c|}{{ AUROC}}         & \multicolumn{1}{c|}{{ AUPR}} \\ \hline
2                     & \multicolumn{1}{c|}{{ 0.68 (0.14)}} & \multicolumn{1}{c|}{{ 0.827 (0.026)}} & { 0.482 (0.056)} & \multicolumn{1}{l|}{{ 0.83 (0.12)}} & \multicolumn{1}{l|}{{ 0.810 (0.032)}} & { 0.488 (0.069)}             \\ \hline
4                     & \multicolumn{1}{c|}{{ 0.87 (0.08)}} & \multicolumn{1}{c|}{{ 0.920 (0.015)}} & { 0.730 (0.055)} & \multicolumn{1}{l|}{{ 0.88 (0.07)}} & \multicolumn{1}{l|}{{ 0.903 (0.022)}} & { 0.736 (0.054)}             \\ \hline
10                    & \multicolumn{1}{c|}{{ 0.87 (0.08)}} & \multicolumn{1}{c|}{{ 0.934 (0.006)}} & { 0.789 (0.020)} & \multicolumn{1}{l|}{{ 0.87 (0.07)}} & \multicolumn{1}{l|}{{ 0.905 (0.019)}} & { 0.755 (0.049)}             \\ \hline
\end{tabular}
}
\end{table}

Table \ref{tab:sim_trunc_pred} further details the performance of VI-LSPM and MCMC-LSPM for different truncation levels $p$ by considering the accuracy, $F_1$ score, and Hamming distance measures; intuitively, the VI-LSPM posterior predictive performance is less accurate than the MCMC-LSPM. Among the truncation levels, $p=4$ naturally performs the best as it fits the correct number of dimensions while $p=\{2, 10\}$ perform slightly worse. The posterior predictive network density and transitivity statistics are overestimated when compared to the observed network for all truncation levels. These results from Table \ref{tab:sim_trunc_pred} are expected as the VI-LSPM has several levels of approximations as detailed in Section \ref{sec:variational}, resulting in inherent bias in the inference. However, the level of bias observed is a worthy trade-off for the faster computational speed of VI-LSPM.

\begin{table}[htb]
\centering
\caption{Performance of V-LSPM and MCMC-LSPM for different truncation levels $p$, assessed through accuracy, $F_1$ score, Hamming distance, and posterior predictive network density and transitivity statistics. Standard deviations are given in brackets. The observed network density for the simulated networks is 0.21 (0.02) while the observed network transitivity is 0.50 (0.02). }
\label{tab:sim_trunc_pred}
\resizebox{\linewidth}{!}{
\begin{tabular}{|c|cc|cc|cc|cc|cc|}
\hline
{ }                      & \multicolumn{2}{c|}{{ Accuracy}}                                         & \multicolumn{2}{c|}{{ $F_1$ score}}                                      & \multicolumn{2}{c|}{{ Hamming distance}}                                 & \multicolumn{2}{c|}{{ Network density}}                                  & \multicolumn{2}{c|}{{ Network transitivity}}                             \\ \cline{2-11} 
\multirow{-2}{*}{{ $p$}} & \multicolumn{1}{c|}{{ VI}}          & { MCMC}        & \multicolumn{1}{c|}{{ VI}}          & { MCMC}        & \multicolumn{1}{c|}{{ VI}}          & { MCMC}        & \multicolumn{1}{c|}{{ VI}}          & { MCMC}        & \multicolumn{1}{c|}{{ VI}}          & { MCMC}        \\ \hline
{ 2}                     & \multicolumn{1}{c|}{{ 0.63 (0.04)}} & { 0.85 (0.03)} & \multicolumn{1}{c|}{{ 0.44 (0.03)}} & { 0.63 (0.08)} & \multicolumn{1}{c|}{{ 0.37 (0.04)}} & { 0.16 (0.03)} & \multicolumn{1}{c|}{{ 0.45 (0.05)}} & { 0.21 (0.02)} & \multicolumn{1}{c|}{{ 0.56 (0.04)}} & { 0.46 (0.04)} \\ \hline
{ 4}                     & \multicolumn{1}{c|}{{ 0.68 (0.03)}} & { 0.89 (0.02)} & \multicolumn{1}{c|}{{ 0.52 (0.02)}} & { 0.75 (0.04)} & \multicolumn{1}{c|}{{ 0.33 (0.03)}} & { 0.11 (0.02)} & \multicolumn{1}{c|}{{ 0.46 (0.05)}} & { 0.21 (0.02)} & \multicolumn{1}{c|}{{ 0.60 (0.05)}} & { 0.50 (0.02)} \\ \hline
{ 10}                    & \multicolumn{1}{c|}{{ 0.63 (0.02)}} & { 0.89 (0.02)} & \multicolumn{1}{c|}{{ 0.51 (0.02)}} & { 0.74 (0.04)} & \multicolumn{1}{c|}{{ 0.38 (0.03)}} & { 0.11 (0.02)} & \multicolumn{1}{c|}{{ 0.56 (0.04)}} & { 0.21 (0.02)} & \multicolumn{1}{c|}{{ 0.72 (0.03)}} & { 0.50 (0.02)} \\ \hline
\end{tabular}
}
\end{table}



\begin{figure}[htb]
    \centering
    \includegraphics[width=\linewidth]{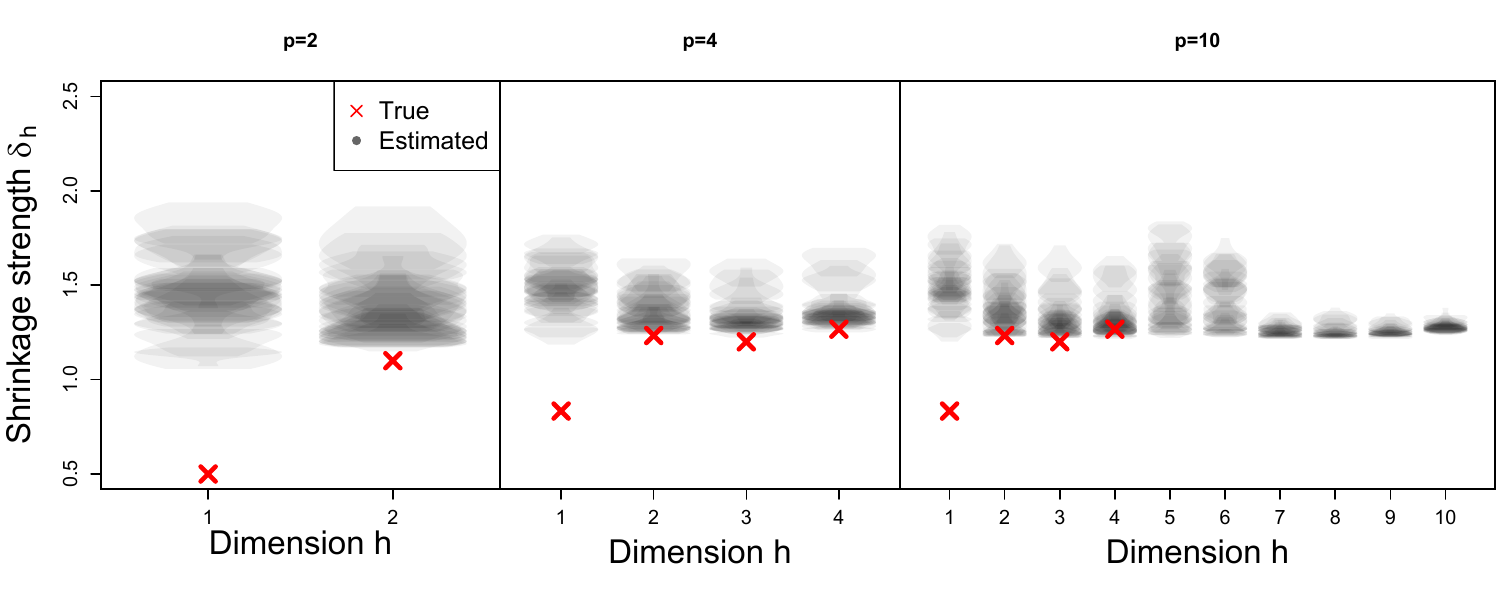}
    \caption{Estimates of the variational posterior shrinkage strength across different truncation levels $p = \{2, 4, 10\}$. Each dimension has 30 violin plots, each representing the 10 point estimates from running VI-LSPM with different starting positions, for the 30 simulated networks. Red crosses are the true parameter values used to simulate the networks. }
    \label{fig:vlspm-sim_trunc_pmd}
\end{figure}

Figure \ref{fig:vlspm-sim_trunc_pmd} shows that the variational posterior shrinkage strength estimates have similar levels of accuracy regardless of the truncation level. When $p > p^* = 4$, Figure \ref{fig:vlspm-sim_trunc_pmd} shows that there is a large increase in the shrinkage strength estimates on the 5th dimension, while, when $p \le p^*$, there is no sudden increase in shrinkage strength. Since a large shrinkage strength means relatively smaller variance of the latent positions on that dimension, this signifies the change from an effective dimension at $p = 4$ to a non-effective dimension at $p = 5$ in the $p > p^* = 4$ case. Although dimensions 7 to 10 tend to have relatively small shrinkage strength compared to dimension 5, note that the precision of the latent positions is a cumulative product of the shrinkage strengths from the previous dimensions. Therefore, the dimensions above 5 will also have small variance and be non-effective. The absence of a large increase in the posterior mean shrinkage strength across dimensions indicates that either the truncation level employed is too low or is equal to the true effective dimension. This behaviour suggests that using a reasonably high truncation level is advisable in order to obtain a clear indication of the number of effective dimensions. However, specifying $p$ to be too large has a negative impact on computational efficiency.

\begin{figure}[htb]
    \centering
    \begin{subfigure}[b]{0.48\linewidth}
         \centering
         \includegraphics[width=\linewidth]{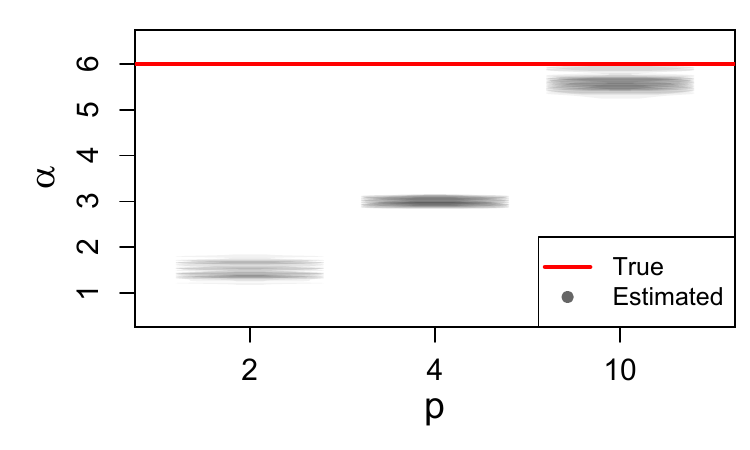}
         \caption{}
         \label{fig:vlspm-trunc_alpha_var}
     \end{subfigure}
     \begin{subfigure}[b]{0.48\linewidth}
         \centering
         \includegraphics[width=\linewidth]{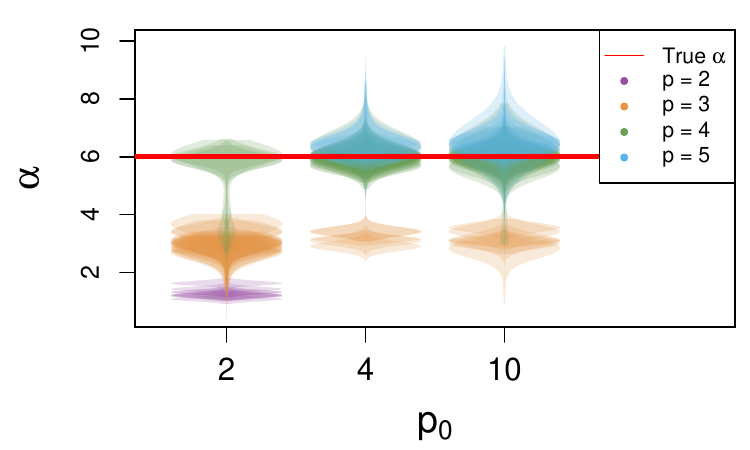}
         \caption{}
         \label{fig:vlspm-trunc_alpha_mcmc}
     \end{subfigure}

    \caption{(a) VI-LSPM estimates of $\alpha$ across different truncation levels $p = \{2, 4, 10\}$. Each dimension has 30 violin plots, each representing the 10 point estimates from running VI-LSPM with different starting positions, for the 30 simulated networks; (b) MCMC-LSPM posterior distributions of $\alpha$ for different initial truncation levels $p_0$ and inferred values of $p$, across 30 simulated networks.}
    \label{fig:vlspm-sim_trunc_alpha}
\end{figure}

Both Figure \ref{fig:vlspm-sim_trunc_pmd} and Figure \ref{fig:vlspm-trunc_alpha_var} show differences between parameters estimates and true values, particularly for $\delta_1$ and $\alpha$, indicating the presence of bias regardless of the truncation level. In contrast, Figure \ref{fig:vlspm-trunc_alpha_mcmc} shows that under the MCMC approach, except when $p < p^*$, there is little bias when estimating $\alpha$. However, it should be noted that the lack of bias observed in Figure \ref{fig:vlspm-trunc_alpha_mcmc} is confounded with the fact that while $p_0$ denotes the initial truncation level in the MCMC approach, an adaptive procedure follows which typically leads to fewer inferred dimensions $p$ and a better fitting model. In summary, while there is a loss in accuracy, this is the trade-off for improved computational cost resulting from using an approximation-based inferential method.

\subsection{Study 2: network size}
\label{ssec:vlspm-size}
Here the focus is on assessing the effect of different network sizes on computational performance and accuracy of the inference. Networks with $n=\{20, 50, 100, 200, 500, 1000\}$ are generated with $p^*=2$, and shrinkage strengths of $\delta_1=0.5$ and $\delta_2=1.1$ giving $\omega_{1}^{-1} = 2$ and $\omega_{2}^{-1} = 1.82$ respectively. Here, $\alpha=3$ is used meaning moderate network density of $\simeq 20\%$. The VI-LSPM with truncation level $p=5$ is fitted and inference on the number of effective dimensions is achieved through inspection of the variational posterior shrinkage strength parameters. Note that, for networks with $n = 1000$, due to the run times involved, the VI-LSPM is run from five sets (rather than 10 sets) of initial positions across the 30 simulated networks, while the MCMC-LSPM is run for only two simulated networks from one set of initial positions. 

As observed in Section \ref{ssec:vlspm-truncation}, Table \ref{tab:sim_size_time} shows notably increased computational speed for the VI-LSPM approach compared to the MCMC-LSPM, where, as in \cite{gwee_2022_a}, 500,000 iterations are used. Table \ref{tab:sim_size_compare} shows that the VI-LSPM demonstrates strong performance as evidenced by the high PC, AUROC, and AUPR values, both when compared with the truth and with the MCMC solutions. As $n$ increases, Table \ref{tab:sim_size_compare} shows increasingly precise PC, AUROC and AUPR values when VI-LSPM is compared against both the truth and the MCMC-LSPM estimates. Intuitively, as shown in Table \ref{tab:sim_size_pred}, the VI-LSPM demonstrates slightly poorer performance than the MCMC approach in terms of accuracy, $F_1$ score, and Hamming distance and shows the VI-LSPM's posterior predicted networks tend to exhibit overestimated network density and transitivity statistics. The bias of the VI-LSPM inference here is similar to that observed in Section \ref{ssec:vlspm-truncation}, which is again the trade-off for having notably faster computational speed. 

\begin{table}[htb]
\centering
\caption{Comparison of VI-LSPM and MCMC-LSPM average model run time in seconds across different network sizes $n$. Standard deviations are given in brackets. *The results for $n=1000$ are based on only five sets of initial positions for VI-LSPM, and only on two networks with one set of initial positions for MCMC-LSPM.}
\label{tab:sim_size_time}
\begin{tabular}{|c|ll|}
\hline
                             & \multicolumn{2}{c|}{Time (sec)}                                                                   \\ \cline{2-3} 
\multirow{-2}{*}{$n$}        & \multicolumn{1}{c|}{VI}                                   & \multicolumn{1}{c|}{MCMC}             \\ \hline
20                           & \multicolumn{1}{l|}{{ 0.6 (0.6)}}     & 304 (15)                              \\ \hline
50                           & \multicolumn{1}{l|}{{ 24 (11)}}       & 481 (4)                               \\ \hline
100                          & \multicolumn{1}{l|}{{ 85 (29)}}       & 1,169 (8)                             \\ \hline
200                          & \multicolumn{1}{l|}{{ 389 (201)}}     & 3,920 (132)                           \\ \hline
{ 500}   & \multicolumn{1}{l|}{{ 1,669 (262)}}   & { 23,264 (2,541)} \\ \hline
{ 1000*} & \multicolumn{1}{l|}{{ 8,419 (1,649)}} & { 105,428 (10,112)}               \\ \hline
\end{tabular}
\end{table}

\begin{table}[htb]
\centering
\caption{Performance of the VI-LSPM across different network sizes $n$, assessed through Procrustes correlation (PC), the area under the receiver operating characteristic curve (AUROC), and the area under the precision recall curve (AUPR) through comparison with the truth and with the MCMC-LSPM solutions. Standard deviations are given in brackets. *The results for $n=1000$ are based on only five sets of initial positions for VI-LSPM, and only on two networks with one set of initial positions for MCMC-LSPM. }
\label{tab:sim_size_compare}
\resizebox{\linewidth}{!}{
\begin{tabular}{|c|ccc|ccc|}
\hline
                             & \multicolumn{3}{c|}{Comparison with truth}                                                                                                                 & \multicolumn{3}{c|}{{ Comparison with MCMC}}                                                                                           \\ \cline{2-7} 
\multirow{-2}{*}{$n$}        & \multicolumn{1}{c|}{PC}                                 & \multicolumn{1}{c|}{AUROC}                                & AUPR                                 & \multicolumn{1}{c|}{{ PC}}          & \multicolumn{1}{c|}{{ AUROC}}         & { AUPR}          \\ \hline
20                           & \multicolumn{1}{c|}{0.78 (0.16)}                        & \multicolumn{1}{c|}{{ 0.920 (0.017)}} & { 0.804 (0.059)} & \multicolumn{1}{c|}{{ 0.77 (0.16)}} & \multicolumn{1}{c|}{{ 0.836 (0.060)}} & { 0.850 (0.073)} \\ \hline
50                           & \multicolumn{1}{c|}{0.93 (0.04)}                        & \multicolumn{1}{c|}{{ 0.904 (0.009)}} & { 0.783 (0.030)} & \multicolumn{1}{c|}{{ 0.93 (0.04)}} & \multicolumn{1}{c|}{{ 0.885 (0.012)}} & { 0.787 (0.041)} \\ \hline
100                          & \multicolumn{1}{c|}{0.95 (0.01)}                        & \multicolumn{1}{c|}{{ 0.904 (0.006)}} & { 0.789 (0.018)} & \multicolumn{1}{c|}{{ 0.95 (0.01)}} & \multicolumn{1}{c|}{{ 0.896 (0.010)}} & { 0.800 (0.025)} \\ \hline
200                          & \multicolumn{1}{c|}{0.96 (0.01)}                        & \multicolumn{1}{c|}{{ 0.904 (0.004)}} & { 0.796 (0.012)} & \multicolumn{1}{c|}{{ 0.96 (0.01)}} & \multicolumn{1}{c|}{{ 0.904 (0.005)}} & { 0.828 (0.025)} \\ \hline
{ 500}   & \multicolumn{1}{c|}{{ 0.97 (0.01)}} & \multicolumn{1}{c|}{{ 0.904 (0.002)}} & { 0.805 (0.005)} & \multicolumn{1}{c|}{{ 0.97 (0.01)}} & \multicolumn{1}{c|}{{ 0.910 (0.005)}} & { 0.865 (0.020)} \\ \hline
{ 1000*} & \multicolumn{1}{c|}{{ 0.97 (0.01)}} & \multicolumn{1}{c|}{{ 0.904 (0.001)}} & { 0.806 (0.007)} & \multicolumn{1}{c|}{{ 0.97 (0.01)}}           & \multicolumn{1}{c|}{{ 0.912 (0.004)}}             & { 0.856 (0.003)}             \\ \hline
\end{tabular}
}
\end{table}

\begin{table}[htb]
\centering
\caption{Performance of V-LSPM and MCMC-LSPM across different network sizes $n$, assessed through accuracy, $F_1$ score, Hamming distance, and posterior predictive network density and transitivity statistics. Standard deviations are given in brackets. The observed network density for the simulated networks is 0.31 (0.03) while the observed network transitivity is 0.58 (0.03). *The results for $n=1000$ are based on only five sets of initial positions for VI-LSPM, and only on two networks with one set of initial positions for MCMC-LSPM.}
\label{tab:sim_size_pred}
\resizebox{\linewidth}{!}{
\begin{tabular}{|c|cc|cc|cc|cc|cc|}
\hline
                             & \multicolumn{2}{c|}{{ Accuracy}}                                         & \multicolumn{2}{c|}{{ $F_1$ score}}                                      & \multicolumn{2}{c|}{{ Hamming distance}}                                 & \multicolumn{2}{c|}{{ Network density}}                                  & \multicolumn{2}{c|}{{ Network transitivity}}                             \\ \cline{2-11} 
\multirow{-2}{*}{$n$}        & \multicolumn{1}{c|}{{ VI}}          & { MCMC}        & \multicolumn{1}{c|}{{ VI}}          & { MCMC}        & \multicolumn{1}{c|}{{ VI}}          & { MCMC}        & \multicolumn{1}{c|}{{ VI}}          & { MCMC}        & \multicolumn{1}{c|}{{ VI}}          & { MCMC}        \\ \hline
20                           & \multicolumn{1}{c|}{{ 0.71 (0.05)}} & { 0.84 (0.04)} & \multicolumn{1}{c|}{{ 0.07 (0.21)}} & { 0.72 (0.06)} & \multicolumn{1}{c|}{{ 0.30 (0.05)}} & { 0.17 (0.04)} & \multicolumn{1}{c|}{{ 0.06 (0.20)}} & { 0.31 (0.06)} & \multicolumn{1}{c|}{{ 0.97 (0.08)}} & { 0.56 (0.08)} \\ \hline
50                           & \multicolumn{1}{c|}{{ 0.66 (0.03)}} & { 0.83 (0.02)} & \multicolumn{1}{c|}{{ 0.62 (0.02)}} & { 0.73 (0.02)} & \multicolumn{1}{c|}{{ 0.34 (0.03)}} & { 0.17 (0.02)} & \multicolumn{1}{c|}{{ 0.60 (0.08)}} & { 0.32 (0.04)} & \multicolumn{1}{c|}{{ 0.74 (0.06)}} & { 0.57 (0.04)} \\ \hline
100                          & \multicolumn{1}{c|}{{ 0.63 (0.02)}} & { 0.83 (0.01)} & \multicolumn{1}{c|}{{ 0.61 (0.02)}} & { 0.73 (0.01)} & \multicolumn{1}{c|}{{ 0.37 (0.02)}} & { 0.17 (0.01)} & \multicolumn{1}{c|}{{ 0.63 (0.05)}} & { 0.32 (0.03)} & \multicolumn{1}{c|}{{ 0.74 (0.04)}} & { 0.58 (0.02)} \\ \hline
200                          & \multicolumn{1}{c|}{{ 0.61 (0.01)}} & { 0.83 (0.01)} & \multicolumn{1}{c|}{{ 0.59 (0.01)}} & { 0.73 (0.01)} & \multicolumn{1}{c|}{{ 0.39 (0.01)}} & { 0.17 (0.01)} & \multicolumn{1}{c|}{{ 0.64 (0.03)}} & { 0.31 (0.02)} & \multicolumn{1}{c|}{{ 0.73 (0.02)}} & { 0.58 (0.01)} \\ \hline
{ 500}   & \multicolumn{1}{c|}{{ 0.61 (0.04)}} & { 0.83 (0.01)} & \multicolumn{1}{c|}{{ 0.57 (0.01)}} & { 0.73 (0.01)} & \multicolumn{1}{c|}{{ 0.39 (0.04)}} & { 0.17 (0.01)} & \multicolumn{1}{c|}{{ 0.61 (0.09)}} & { 0.32 (0.02)} & \multicolumn{1}{c|}{{ 0.68 (0.07)}} & { 0.59 (0.02)} \\ \hline
{ 1000*} & \multicolumn{1}{c|}{{ 0.66 (0.01)}} & { 0.83 (0.01)}           & \multicolumn{1}{c|}{{ 0.58 (0.01)}} & { 0.74 (0.01)}           & \multicolumn{1}{c|}{{ 0.34 (0.01)}} & { 0.17 (0.01)}           & \multicolumn{1}{c|}{{ 0.50 (0.02)}} & { 0.33 (0.03)}           & \multicolumn{1}{c|}{{ 0.57 (0.01)}} & { 0.60 (0.02)}           \\ \hline
\end{tabular}
}
\end{table}

\begin{figure}[htb]
    \centering
    \includegraphics[width=\linewidth]{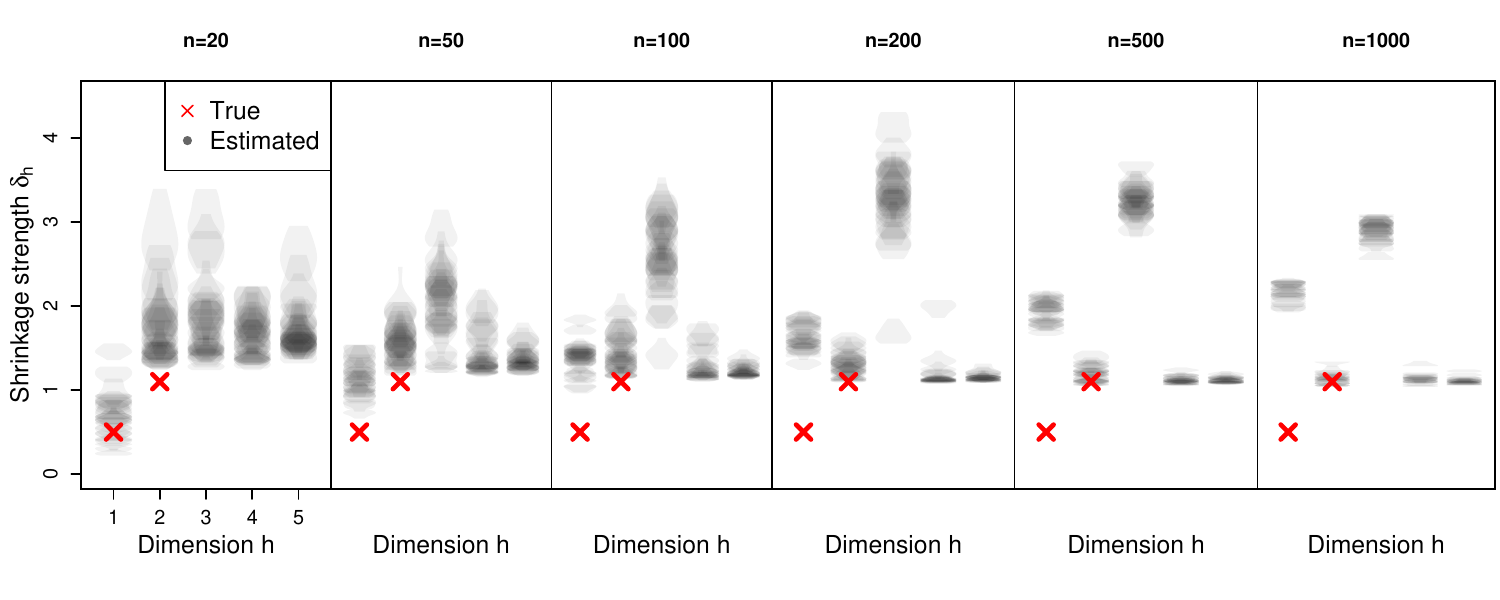}
    \caption{Estimates of the variational posterior shrinkage strength across networks of different sizes: $n=\{20, 50, 100, 200, 500, 1000\}$. Each dimension has 30 violin plots, each representing the 10 point estimates from running VI-LSPM with different starting positions, for the 30 simulated networks. Red crosses are the true parameter values used to simulate the networks. For $n=1000$, the violin plots represent only five point estimates for each of the 30 simulated networks. }
    \label{fig:vlspm-sim_size_pmd}
\end{figure}

Figure \ref{fig:vlspm-sim_size_pmd} shows that, with increasing numbers of nodes, the variational posterior shrinkage strength for dimension 3 becomes increasingly large in value, signifying the increasingly clearer change from an effective dimension at $p = 2$ to a non-effective dimension at $p = 3$. 

\section{Illustrative network data examples}
\label{sec:vlspm-appl}
This section investigates the performance of VI-LSPM on illustrative network data examples. In addition to the VI-LSPM and the MCMC-LSPM, also the variational-based LPM (VI-LPM) and the MCMC-based LPM (MCMC-LPM) are included in the comparison. The VI-LPM is implemented through the R package \texttt{lvm4net} \citep{gollini_2016_joint}, while the MCMC-LPM is implemented using the R package \texttt{latentnet} \citep{krivitsky_2008_fitting, latentnetR}.

\subsection{Cat brain connectivity binary network data}
\label{ssec:vlspm-cat}
The VI-LSPM is used to analyse a cat brain connectivity network which encompasses 65 distinct cortex regions, represented as nodes, and 1139 interregional macroscopic axonal projections, represented as edges \citep{scannell_1995_analysis, de_2013_rich}. The regions within the cortex are grouped into four primary categories: visual (18 areas), auditory (10 areas), somatomotor (18 areas), and frontolimbic (19 areas). These groupings stem from neurophysiological data detailing the specific function of each brain region. Here, the network is viewed as a binary directed network. A value of $y_{i,j} = 1$ suggests a connection between regions $i$ and $j$, while $y_{i,j} = 0$ denotes the absence of such a connection. The network density is $27.37\%$ and transitivity is 0.52.

\begin{figure}[htb!]
     \centering
     \begin{subfigure}[b]{0.8\linewidth}
         \centering
         \includegraphics[width=0.7\linewidth]{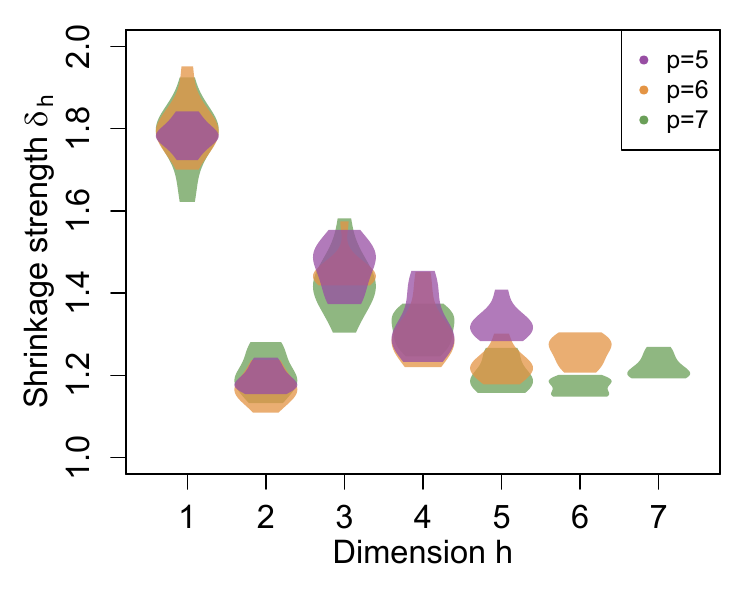}
         \caption{}
         \label{fig:vlspm-cat_pmd}
     \end{subfigure}
     \begin{subfigure}[b]{0.495\linewidth}
         \centering
         \includegraphics[width=\linewidth]{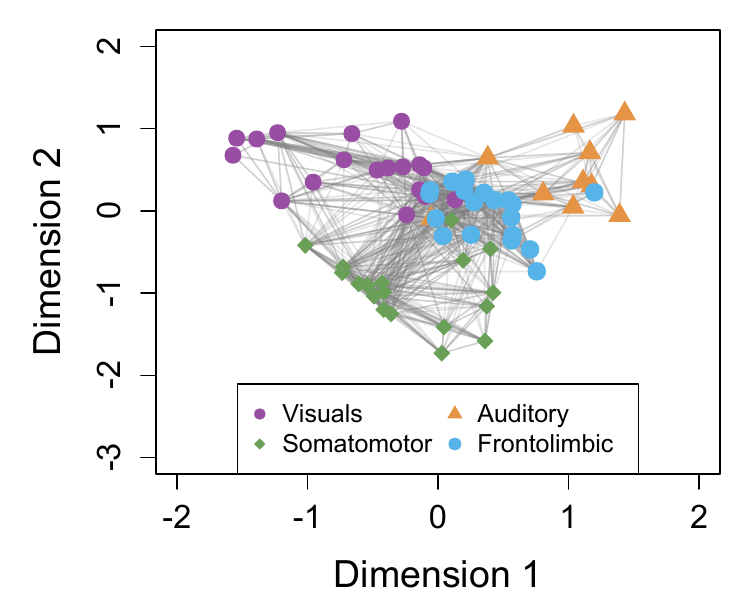}
         \caption{}
         \label{fig:vlspm-cat_vlspm_pos}
     \end{subfigure}
     \begin{subfigure}[b]{0.495\linewidth}
         \centering
         \includegraphics[width=\linewidth]{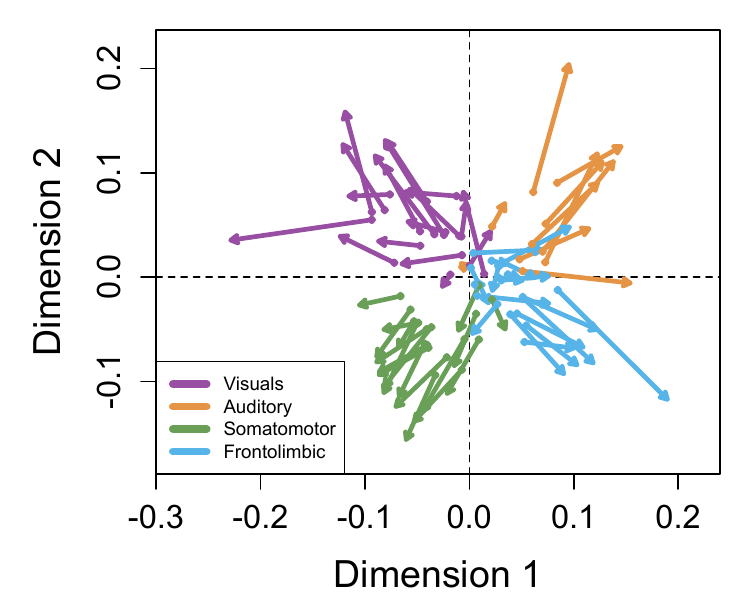}
         \caption{}
         \label{fig:vlspm-cat_vlspm_protest}
     \end{subfigure}
        \caption{Cat brain connectivity data. (a) The VI-LSPM estimates of the variational posterior shrinkage strength parameters from multiple runs of different starting values coloured according to different truncation levels, (b) the inferred 2-dimensional variational posterior latent positions under one of the VI-LSPM runs, and 
        (c) the Procrustes errors between the VI-LSPM variational posterior latent positions (arrow tail) and the MCMC-LSPM posterior mean latent positions on the first two dimensions (arrow head).}
        \label{fig:vlspm-cat}
\end{figure}

Fitting the MCMC-LSPM took an average of 10 ($\pm 1$) minutes (from 10 different starting values) with 500,000 iterations, while the variational VI-LSPM (from 10 different starting values) with $p=5$ took on average 16 ($\pm 4$) seconds and required 23 ($\pm 3$) iterations (standard deviations are given in brackets). Figure \ref{fig:vlspm-cat_pmd} reports the estimated shrinkage strength parameters and suggests 2 dimensions as the number of effective dimensions; the MCMC approach suggests 3 via the posterior mode of the number of dimension. Figure \ref{fig:vlspm-cat_pmd} also shows that the inference on the number of dimensions is insensitive to the dimension truncation level. The Procrustes correlation is assessed between the VI-LSPM latent positions and the 2-dimensional latent positions obtained from the three other inferential methods: comparing with MCMC-LSPM gives PC = 0.69 ($\pm 0.01$), with MCMC-LPM gives PC = 0.66 ($\pm 0.01$), and with VI-LPM gives PC = 0.67 ($\pm 0.01$). 

In terms of model fit, the AUROC and AUPR from the VI-LSPM are 0.891 ($\pm 0.006$) and 0.733 ($\pm 0.012$) respectively, suggesting good model fit. Posterior predictive checks result in accuracy = 0.64 ($\pm 0.01$), $F_1$ score = 0.56 ($\pm 0.01$), Hamming distance = 0.36 ($\pm 0.01$), network density = 0.56 ($\pm 0.01$), and network transitivity = 0.67 ($\pm 0.01$). Intuitively, the predicted network statistics are overestimated, given the approximate nature of the VI-LSPM's inferential procedure. However, as the run times indicate, these biases are in return for increased computational efficiency.


Figure \ref{fig:vlspm-cat_vlspm_pos} shows the variational posterior latent positions under the VI-LSPM. Nodes from the same brain region are closely located with nodes from the visuals and somatomotor regions lying below zero on the first dimension while auditory and frontolimbic nodes mostly lie above zero. Similarly on dimension 2, nodes from the somatomotor region tend to lie below zero. Figure \ref{fig:vlspm-cat_vlspm_protest} shows the Procrustes errors between the inferred VI-LSPM variational posterior latent positions and the MCMC-LSPM posterior mean latent positions on the first two dimensions, indicating the overall structure of the configurations is very similar with the VI-LSPM positions being less dispersed than their MCMC counterparts.


\subsection{Worm nervous system binary network data}
\label{ssec:vlspm-worm}
This binary directed network comprises $n=272$ nodes of neurons derived from the nervous system of the adult male \emph{Caenorhabditis elegans} worm. Before analysis, three nodes without edges were removed. Each of the nodes belongs to one of 5 cell types: muscle (90 nodes), sensory (72 nodes), interneuron (67 nodes), motorneuron (38 nodes), and other (2 nodes). Each of the 4451 edges signifies either a chemical or electrical interaction between a pair of nodes. The data were compiled from serial electron micrograph sections, as documented by \cite{jarrell_2012_the}. The network density is 6.09\% and transitivity is 0.29.

\begin{figure}[htb!]
    \centering
     \begin{subfigure}[b]{0.48\linewidth}
         \centering
         \includegraphics[width=\linewidth]{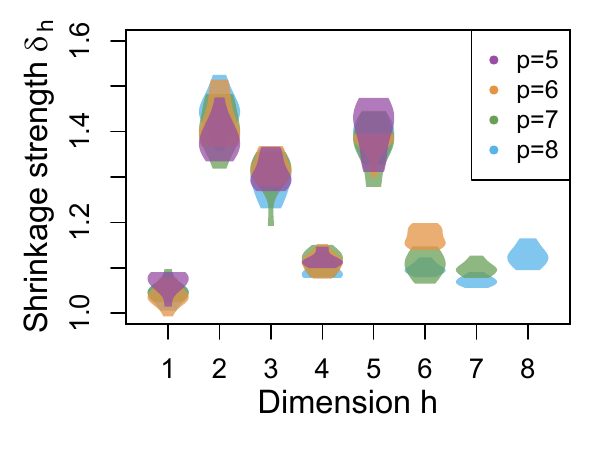}
         \caption{}
         \label{fig:vlspm-worm_pmd}
     \end{subfigure}
     \begin{subfigure}[b]{0.48\linewidth}
         \centering
         \includegraphics[width=\linewidth]{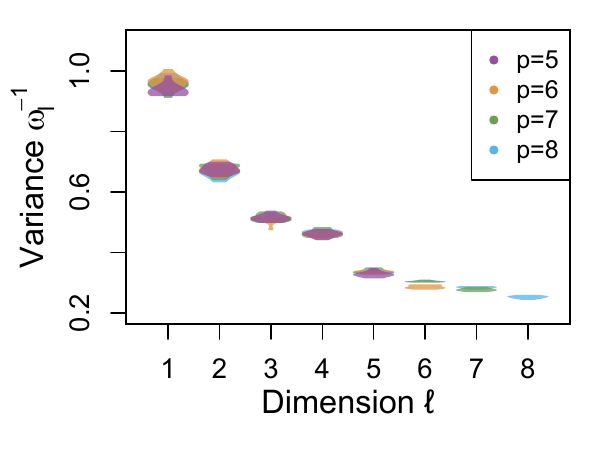}
         \caption{}
         \label{fig:vlspm-worm_pmv}
     \end{subfigure}
    \caption{For the worm nervous system network, the VI-LSPM point estimates of the (a) variational posterior shrinkage strength parameters and (b) variational posterior variance parameters.}
    \label{fig:vlspm-worm}
\end{figure}

Model fitting took an average of 225 ($\pm 28$) minutes (from 10 different starting values) via the MCMC approach with 1,000,000 iterations. On the other hand, the VI-LSPM (from 10 different starting values) took on average 8 ($\pm 1$) minutes with 43 ($\pm 2$) iterations (standard deviations given in brackets). Figure \ref{fig:vlspm-worm} indicates 4 effective dimensions, while the MCMC approach indicated 3 via the posterior mode of the number of dimension. The Procrustes correlation is assessed between the latent positions on the first 3-dimensions between the VI-LSPM and the other methods considered for comparison: with MCMC-LSPM, PC = 0.752 ($\pm 0.004$), with MCMC-LPM, PC = 0.671 ($\pm 0.004$) and with the VI-LPM, PC = 0.782 ($\pm 0.005$). 
Also, the AUROC and AUPR metrics for the VI-LSPM are 0.853 ($\pm 0.002$) and 0.257 ($\pm 0.008$) respectively, suggesting good model fit. While the AUPR value is low, it is higher than the baseline corresponding to the network density, still indicating good model fit in terms of identification of edges. Posterior predictive checks of the VI-LSPM resulted in accuracy = 0.73 ($\pm 0.01$), $F_1$ score = 0.24 ($\pm 0.01$), Hamming distance = 0.27 ($\pm 0.01$), network density = 0.29 ($\pm 0.01$), and network transitivity = 0.55 ($\pm 0.01$). Similar to Section \ref{ssec:vlspm-cat}, the predicted network statistics tend to be overestimated. Again, however, as the run times indicate, these biases are in exchange of improved computational efficiency.

\begin{figure}[htb!]
    \centering
    \includegraphics[width=\linewidth]{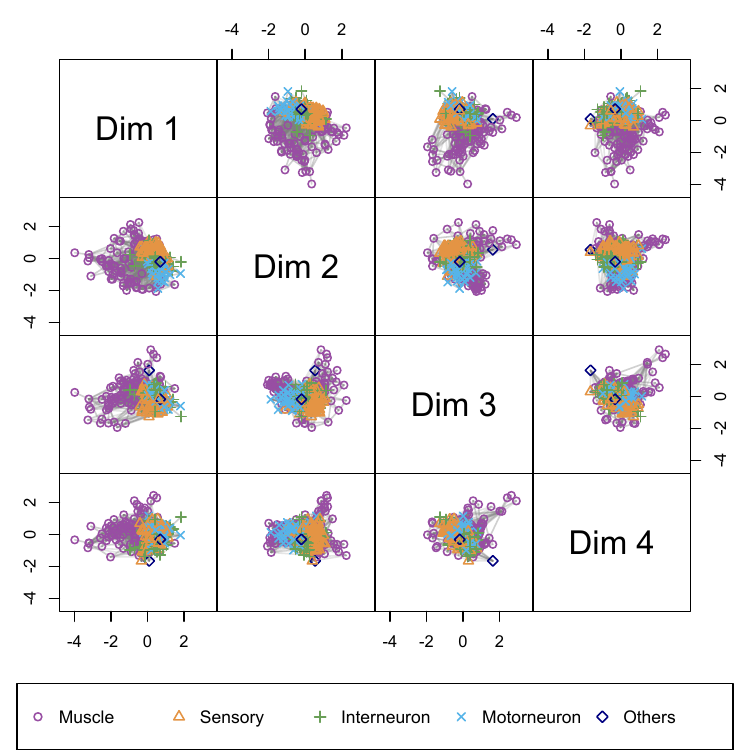}
    \caption{For the worm nervous system network, the latent position estimates on the first 4 dimensions under the VI-LSPM coloured by the cell types.}
    \label{fig:vlspm-worm_vlspm_pairs}
\end{figure}

Figure \ref{fig:vlspm-worm_vlspm_pairs} shows the variational posterior latent positions' estimates on the first 4 dimensions.  Nodes from the muscle cell type have the largest variance across dimensions but tend to be negative on the first dimension. Nodes belonging to the rest of the cell types are closely located to each other in the latent space. On dimension 2, the sensory cells tend to be positive, while the motorneuron cells are located below zero. 

As shown by the PC values, while there is good correspondence between the latent configurations under the MCMC and variational approaches, under the VI-LSPM, the latent positions struggle to move from the initial configuration, initialised as detailed in Section \ref{ssec:vlspm-update}, with a PC between the initial and estimated configurations of 0.99 $\pm <0.01$. The MCMC approach instead manages to move away from the initial configuration with PC between the initial and posterior mean configurations of 0.56 $\pm 0.06$. Considering alternative initialisation strategies for the VI-LSPM may help here. Also, the VI-LSPM does not shrink the variance of the latent positions on the non-effective dimensions as strongly as its MCMC counterpart. However, inference on the number of effective dimensions was still clear in the case of the network considered here. For reference, the initial positions and the MCMC-LSPM posterior mean positions are given in Appendix \ref{app:vlspm-pos}.



\section{Discussion} \label{sec:vlspm-disc}
In this work a variational inference approach for the LSPM has been developed and has shown an improvement in computational cost when compared to the MCMC approach. Although the variational approach is approximate, the VI-LSPM demonstrated good accuracy when inferring the number of effective latent dimensions. Moreover, for the VI-LSPM in the simulation studies, there was good correspondence between the estimated and true latent positions, despite some bias in some parameter estimates. Goodness of fit metrics via the AUROC and AUPR also indicated satisfactory performance in both simulation studies and applications. While in-sample metrics are shown to be satisfactory, assessment via out-of-sample metrics such as those considered in \cite{li_2020_network} could be helpful in further validation.

Performance of the VI-LSPM was observed to be poor in networks that were very sparse or very dense, as in such cases $\alpha$ dominates the logistic function in a way that the distances between the latent positions have little influence on the edge formation process. This is especially problematic when the truncation level is different from the true number of effective dimensions, since biases are further incorporated into the estimate of $\alpha$. The MCMC-LSPM addresses this problem via an adaptive sampling procedure which automatically shrinks or grows the number of latent dimensions as required, therefore reducing parameter bias. However, as the VI-LSPM does not sample from the true posterior distribution, a way to perform such an adaptive sampling procedure is less intuitive. To obviate this problem, a different shrinkage prior may be more suitable such as employing the Indian buffet process \citep{knowles_2011_nonparametric, rokov_2016_fast}. 

While the main focus of this work was to develop a computationally efficient approach to inferring the LSPM, the proposed methodology could be easily adapted to a range of network data types, much like its MCMC counterpart. For example, developing a variational approach to inference for the LSPM for networks with count edges as in \cite{gwee_2022_a} or to perform clustering of nodes using the LSPCM as in \cite{gwee_2023_modelbased} would be natural extensions.


\section*{Acknowledgments}
This publication has emanated from research conducted with the financial support of Science
Foundation Ireland under grant number 18/CRT/6049. For the purpose of Open Access, the
author has applied a CC BY public copyright licence to any Author Accepted Manuscript
version arising from this submission. 

The authors are grateful for the discussions with Dr Isabella Gollini that assisted with this work.








\clearpage

\appendix
\section{Derivation of the KL divergence} \label{app:vlspm-deriveKL}
The derivation of the KL divergence is derived with the help of the matrix cookbook \citep{matrix_cookbook}. By definition, the KL divergence between the variational posterior $\mathbb{Q}(\alpha, \mathbf{Z}, \bm{\delta})$ and the true posterior $\mathbb{P}(\alpha, \mathbf{Z}, \bm{\delta}  \mid \mathbf{Y})$ is:

$$
\begin{aligned}
&\mathrm{KL}[\mathbb{Q}(\alpha, \mathbf{Z}, \bm{\delta}) \| \mathbb{P}(\alpha, \mathbf{Z}, \bm{\delta} \mid \mathbf{Y})] \\
& =-\int \mathbb{Q}(\alpha, \mathbf{Z}, \bm{\delta}) \log \frac{\mathbb{P}(\alpha, \mathbf{Z}, \bm{\delta} \mid \mathbf{Y})}{\mathbb{Q}(\alpha, \mathbf{Z}, \bm{\delta})} d(\alpha, \mathbf{Z}, \bm{\delta}) \\
& =-\int \mathbb{Q}(\alpha, \mathbf{Z}, \bm{\delta}) \log \frac{\mathbb{P}(\mathbf{Y}, \alpha, \mathbf{Z}, \bm{\delta})}{\mathbb{P}(\mathbf{Y})\mathbb{Q}(\alpha, \mathbf{Z}, \bm{\delta})} d(\alpha, \mathbf{Z}, \bm{\delta}) \\
&=-\int \mathbb{Q}(\alpha, \mathbf{Z}, \bm{\delta}) \log \frac{\mathbb{P}(\mathbf{Y}, \alpha, \mathbf{Z}, \bm{\delta})}{\mathbb{Q}(\alpha, \mathbf{Z}, \bm{\delta})} d(\alpha, \mathbf{Z}, \bm{\delta})  + \int \mathbb{Q}(\alpha, \mathbf{Z}, \bm{\delta}) \log \mathbb{P}(\mathbf{Y}) d(\alpha, \mathbf{Z}, \bm{\delta}) \\
&=-\int \mathbb{Q}(\alpha, \mathbf{Z}, \bm{\delta}) \log \frac{\mathbb{P}(\mathbf{Y}, \alpha, \mathbf{Z}, \bm{\delta})}{\mathbb{Q}(\alpha, \mathbf{Z}, \bm{\delta})} d(\alpha, \mathbf{Z}, \bm{\delta}) +\log \mathbb{P}(\mathbf{Y}) \\
&=-\int \mathbb{Q}(\alpha, \mathbf{Z}, \bm{\delta}) \log \frac{\mathbb{P}(\mathbf{Y}\mid\alpha, \mathbf{Z}) \mathbb{P}(\alpha) \prod_{i=1}^n \mathbb{P}(\mathbf{z}_i)\mathbb{P}(\delta_1) \prod_{h=2}^p \mathbb{P}(\delta_h) }{\mathbb{Q}(\alpha) \prod_{i=1}^n \mathbb{Q}(\mathbf{z}_i)\mathbb{Q}(\delta_1) \prod_{h=2}^p \mathbb{Q}(\delta_h)} d(\alpha, \mathbf{Z}, \bm{\delta})  +\log \mathbb{P}(\mathbf{Y}) \\
&=-\int \mathbb{Q}(\alpha, \mathbf{Z}, \bm{\delta}) \log \mathbb{P}(\mathbf{Y}\mid\alpha, \mathbf{Z}) d(\alpha, \mathbf{Z}, \bm{\delta}) +\log \mathbb{P}(\mathbf{Y})  \\
& \qquad -\int \mathbb{Q}(\alpha, \mathbf{Z}, \bm{\delta}) \log \mathbb{P}(\alpha) d(\alpha, \mathbf{Z}, \bm{\delta}) + \int \mathbb{Q}(\alpha, \mathbf{Z}, \bm{\delta}) \log \mathbb{Q}(\alpha) d(\alpha, \mathbf{Z}, \bm{\delta}) \\
& \qquad -\int \mathbb{Q}(\alpha, \mathbf{Z}, \bm{\delta}) \log \prod_{i=1}^n \mathbb{P}(\mathbf{z}_i) d(\alpha, \mathbf{Z}, \bm{\delta}) + \int \mathbb{Q}(\alpha, \mathbf{Z}, \bm{\delta}) \log \prod_{i=1}^n \mathbb{Q}(\mathbf{z}_i) d(\alpha, \mathbf{Z}, \bm{\delta}) \\
& \qquad -\int \mathbb{Q}(\alpha, \mathbf{Z}, \bm{\delta}) \log \mathbb{P}(\delta_1) d(\alpha, \mathbf{Z}, \bm{\delta}) + \int \mathbb{Q}(\alpha, \mathbf{Z}, \bm{\delta}) \log \mathbb{Q}(\delta_1) d(\alpha, \mathbf{Z}, \bm{\delta}) \\
& \qquad -\int \mathbb{Q}(\alpha, \mathbf{Z}, \bm{\delta}) \log \prod_{h=2}^p \mathbb{P}(\delta_h) d(\alpha, \mathbf{Z}, \bm{\delta}) + \int \mathbb{Q}(\alpha, \mathbf{Z}, \bm{\delta}) \log \prod_{h=2}^p \mathbb{Q}(\delta_h) d(\alpha, \mathbf{Z}, \bm{\delta}) \\
&= -\mathbb{E}_{\mathbb{Q}(\mathbf{Z}, \alpha \mid \mathbf{Y})}[\log (\mathbb{P}(\mathbf{Y} \mid \mathbf{Z}, \alpha))] +\log \mathbb{P}(\mathbf{Y})\\
& \qquad - \mathbb{E}_{\mathbb{Q}}[\log \mathbb{P}(\alpha)] + \mathbb{E}_{\mathbb{Q}}[\log \mathbb{Q}(\alpha)]  - \sum_{i=1}^{n} \mathbb{E}_{\mathbb{Q}}\left[\log \mathbb{P}\left(\mathbf{z}_i \right)\right]  + \sum_{i=1}^{n}  \mathbb{E}_{\mathbb{Q}}\left[\log \mathbb{Q}\left(\mathbf{z}_i \right)\right]   \\
& \qquad - \mathbb{E}_{\mathbb{Q}}\left[\log \mathbb{P}\left(\delta_1\right)\right] +\mathbb{E}_{\mathbb{Q}}\left[\log \mathbb{Q}\left(\delta_1\right)\right]  - \sum_{h=2}^p \mathbb{E}_{\mathbb{Q}}\left[\log \mathbb{P}\left(\delta_h\right)\right] + \sum_{h=2}^p \mathbb{E}_{\mathbb{Q}}\left[\log \mathbb{Q}\left(\delta_h\right)\right] \\
& = -\mathbb{E}_{\mathbb{Q}(\mathbf{Z}, \alpha \mid \mathbf{Y})}\left[\log (\mathbb{P}(\mathbf{Y} \mid \mathbf{Z}, \alpha)\right] +\log \mathbb{P}(\mathbf{Y}) + \operatorname{KL}\left[\mathbb{Q}(\alpha) \| \mathbb{P}(\alpha)\right]  \\
&  \qquad +\sum_{i=1}^{n} \operatorname{KL}\left[\mathbb{Q}\left(\mathbf{z}_{i}\right) \| \mathbb{P}\left(\mathbf{z}_{i}\right)\right] +\operatorname{KL}\left[\mathbb{Q}(\delta_1) \| \mathbb{P}(\delta_1)\right]+\sum_{h=2}^p \operatorname{KL}\left[\mathbb{Q}\left(\delta_{h}\right) \| \mathbb{P}\left(\delta_{h}\right)\right] \\
\end{aligned}
$$

\noindent

Here follows the derivation of $\operatorname{KL}\left[\mathbb{Q}(\alpha) \| \mathbb{P}(\alpha)\right]$:
$$
\begin{aligned}
\mathbb{E}_{\mathbb{{\mathbb{Q}}}}[\log {\mathbb{Q}}(\alpha)]&=\mathbb{E}_{\mathbb{{\mathbb{Q}}}}\left[-\frac{1}{2} \log \left(2 \pi \tilde{\sigma}_\alpha^2\right)-\frac{(\alpha-\tilde{\mu}_\alpha)^2}{2\tilde{\sigma}_\alpha^2}\right] \\ 
&=\mathbb{E}_{\mathbb{{\mathbb{Q}}}}\left[-\frac{1}{2} \log \left(2 \pi \tilde{\sigma}_\alpha^2\right)-\frac{(\alpha^2-2\alpha\tilde{\mu}_\alpha+\tilde{\mu}_\alpha^2)}{2\tilde{\sigma}_\alpha^2}\right] \\
&=-\frac{1}{2} \log \left(2 \pi \tilde{\sigma}_\alpha^2\right)-\frac{\mathbb{E}_{\mathbb{{\mathbb{Q}}}}\left[\alpha^2\right]-2\mathbb{E}_{\mathbb{{\mathbb{Q}}}}\left[\alpha\right]\tilde{\mu}_\alpha+\tilde{\mu}_\alpha^2}{2\tilde{\sigma}_\alpha^2} \\
&=-\frac{1}{2} \log \left(2 \pi \tilde{\sigma}_\alpha^2\right)-\frac{\tilde{\mu}_\alpha^2+\tilde{\sigma}_\alpha^2-2\tilde{\mu}_\alpha\tilde{\mu}_\alpha+\tilde{\mu}_\alpha^2}{2\tilde{\sigma}_\alpha^2} \\
&=-\frac{1}{2} \log \left(2 \pi \tilde{\sigma}_\alpha^2\right)-\frac{1}{2} \\
\mathbb{E}_{\mathbb{{\mathbb{Q}}}}[\log {\mathbb{P}}(\alpha)]&=\mathbb{E}_{\mathbb{{\mathbb{Q}}}}\left[-\frac{1}{2} \log \left(2 \pi \sigma_\alpha^2\right)-\frac{(\alpha-\mu_\alpha)^2}{2\sigma_\alpha^2}\right] \\
&=\mathbb{E}_{\mathbb{{\mathbb{Q}}}}\left[-\frac{1}{2} \log \left(2 \pi \sigma_\alpha^2\right)-\frac{(\alpha^2-2\alpha\mu_\alpha+\mu_\alpha^2)}{2\sigma_\alpha^2}\right] \\
&=-\frac{1}{2} \log \left(2 \pi \sigma_\alpha^2\right)-\frac{\mathbb{E}_{\mathbb{{\mathbb{Q}}}}\left[\alpha^2\right]-2\mathbb{E}_{\mathbb{{\mathbb{Q}}}}\left[\alpha\right]\mu_\alpha+\mu_\alpha^2}{2\sigma_\alpha^2} \\
&=-\frac{1}{2} \log \left(2 \pi \sigma_\alpha^2\right)-\frac{\tilde{\mu}_\alpha^2+\tilde{\sigma}_\alpha^2-2\tilde{\mu}_\alpha\mu_\alpha+\mu_\alpha^2}{2\sigma_\alpha^2} \\
\end{aligned} 
$$
$$
\begin{aligned}
\operatorname{KL} \left[\mathbb{Q}(\alpha) \| \mathbb{P}(\alpha)\right] & =\mathbb{E}_\mathbb{Q}\left[\log {\mathbb{Q}}(\alpha)\right] -\mathbb{E}_\mathbb{Q} \left[\log {\mathbb{P}}(\alpha)\right] \\
& = \left\{ -\frac{1}{2} \log \left(2 \pi \tilde{\sigma}_\alpha^2\right)-\frac{1}{2} \right\}  -\left\{ -\frac{1}{2} \log \left(2 \pi \sigma_\alpha^2\right)-\frac{\tilde{\mu}_\alpha^2+\tilde{\sigma}_\alpha^2-2\tilde{\mu}_\alpha\mu_\alpha+\mu_\alpha^2}{2\sigma_\alpha^2} \right\} \\
& = -\frac{1}{2} \log \left( \frac{\tilde{\sigma}_\alpha^2}{\sigma_\alpha^2 }\right) -\frac{1}{2} +\frac{\tilde{\sigma}_\alpha^2+(\tilde{\mu}_\alpha^2-2\tilde{\mu}_\alpha\mu_\alpha+\mu_\alpha^2)}{2\sigma_\alpha^2}  \\
& = -\frac{1}{2} \log \left( \frac{\tilde{\sigma}_\alpha^2}{\sigma_\alpha^2 }\right)-\frac{1}{2} +\frac{\tilde{\sigma}_\alpha^2+(\tilde{\mu}_\alpha-\mu_\alpha)^2}{2\sigma_\alpha^2}  \\
& = -\frac{1}{2} \log \left( \frac{\tilde{\sigma}_\alpha^2}{\sigma_\alpha^2 }\right)-\frac{1}{2} +\frac{\tilde{\sigma}_\alpha^2}{2\sigma_\alpha^2} +\frac{(\tilde{\mu}_\alpha-\mu_\alpha)^2}{2\sigma_\alpha^2}   \\
\end{aligned} 
$$

\noindent
Here follows the derivation of $\operatorname{KL}\left[\mathbb{Q}\left(\mathbf{z}_{i}\right) \| \mathbb{P}\left(\mathbf{z}_{i}\right)\right] $:
$$
\begin{aligned}
\mathbb{E}_{\mathbb{{\mathbb{Q}}}}\left[\log {\mathbb{Q}}\left(\mathbf{z}_i \mid \mathbf{\Omega}\right)\right]&=\mathbb{E}_{\mathbb{{\mathbb{Q}}}}\left[\log \left\{\left(\frac{1}{2\pi}\right)^{p/2} \operatorname{det}(\tilde{\mathbf{\Omega}})^{1/2} \exp \left[-\frac{1}{2}\left(\mathbf{z}_i-\tilde{\mathbf{z}}_i\right)^{\top} \tilde{\mathbf{\Omega}}\left(\mathbf{z}_i-\tilde{\mathbf{z}}_i\right)\right]\right\}\right] \\
&=\mathbb{E}_{\mathbb{{\mathbb{Q}}}}\left[-\frac{p}{2}\log(2\pi) + \frac{1}{2}\log \operatorname{det}(\tilde{\mathbf{\Omega}}) -\frac{1}{2}\left(\mathbf{z}_i-\tilde{\mathbf{z}}_i\right)^{\top} \tilde{\mathbf{\Omega}}\left(\mathbf{z}_i-\tilde{\mathbf{z}}_i\right)\right] \\
&=-\frac{p}{2}\log(2\pi) + \frac{1}{2}\log \operatorname{det}(\tilde{\mathbf{\Omega}}) -\frac{1}{2}\mathbb{E}_{\mathbb{{\mathbb{Q}}}}\left[\left(\mathbf{z}_i-\tilde{\mathbf{z}}_i\right)^{\top} \tilde{\mathbf{\Omega}}\left(\mathbf{z}_i-\tilde{\mathbf{z}}_i\right)\right] \\
&=-\frac{p}{2}\log(2\pi) + \frac{1}{2}\log \operatorname{det}(\tilde{\mathbf{\Omega}}) -\frac{1}{2}\mathbb{E}_{\mathbb{{\mathbb{Q}}}}\left[\mathbf{z}_i^{\top} 	\tilde{\mathbf{\Omega}} \mathbf{z}_i - \mathbf{z}_i^{\top} \tilde{\mathbf{\Omega}} \tilde{\mathbf{z}}_i - \tilde{\mathbf{z}}_i^{\top} 	\tilde{\mathbf{\Omega}} \mathbf{z}_i + \tilde{\mathbf{z}}_i^{\top} \tilde{\mathbf{\Omega}} \tilde{\mathbf{z}}_i \right] \\
&=-\frac{p}{2}\log(2\pi) + \frac{1}{2}\log \operatorname{det}(\tilde{\mathbf{\Omega}}) \\
& \qquad -\frac{1}{2} \left\{\mathbb{E}_{\mathbb{{\mathbb{Q}}}}\left[\mathbf{z}_i^{\top}  \tilde{\mathbf{\Omega}} \mathbf{z}_i \right] -\mathbb{E}_{\mathbb{{\mathbb{Q}}}}\left[ \mathbf{z}_i^{\top} \right] \tilde{\mathbf{\Omega}} \tilde{\mathbf{z}}_i - \tilde{\mathbf{z}}_i^{\top} \tilde{\mathbf{\Omega}} \mathbb{E}_{\mathbb{{\mathbb{Q}}}}\left[\mathbf{z}_i\right] + \tilde{\mathbf{z}}_i^{\top} \tilde{\mathbf{\Omega}} \tilde{\mathbf{z}}_i \right\} \\
&=-\frac{p}{2}\log(2\pi) + \frac{1}{2}\log \operatorname{det}(\tilde{\mathbf{\Omega}}) \\
& \qquad -\frac{1}{2} \left\{\tilde{\mathbf{z}}_i^{\top} \tilde{\mathbf{\Omega}} \tilde{\mathbf{z}}_i + \operatorname{tr}(\tilde{\mathbf{\Omega}} \tilde{\mathbf{\Omega}}^{-1}) -\tilde{\mathbf{z}}_i^{\top} \tilde{\mathbf{\Omega}} \tilde{\mathbf{z}}_i - \tilde{\mathbf{z}}_i^{\top} \tilde{\mathbf{\Omega}} \tilde{\mathbf{z}}_i + \tilde{\mathbf{z}}_i^{\top} \tilde{\mathbf{\Omega}} \tilde{\mathbf{z}}_i \right\} \\
&=-\frac{p}{2}\log(2\pi) + \frac{1}{2}\log \operatorname{det}(\tilde{\mathbf{\Omega}}) -\frac{1}{2} \operatorname{tr}(\tilde{\mathbf{\Omega}} \tilde{\mathbf{\Omega}}^{-1})  \\
&=-\frac{p}{2}\log(2\pi) + \frac{1}{2}\log \operatorname{det}(\tilde{\mathbf{\Omega}}) -\frac{p}{2}  \\
\mathbb{E}_{\mathbb{{\mathbb{Q}}}}\left[\log {\mathbb{P}}\left(\mathbf{z}_i \mid \mathbf{\Omega}\right)\right] &=\mathbb{E}_{\mathbb{{\mathbb{Q}}}}\left[\log \left\{ \left(\frac{1}{2 \pi}\right)^{p / 2} \operatorname{det}(\mathbf{\Omega})^{1 / 2} \exp \left[-\frac{1}{2}\left(\mathbf{z}_i-0\right)^{\top} \mathbf{\Omega}\left(\mathbf{z}_i-0\right)\right]\right\}\right] \\
& =\mathbb{E}_{\mathbb{{\mathbb{Q}}}}\left[\sum_{i=1}^n \left\{ -\frac{p}{2} \log(2\pi) +\frac{1}{2}\log\operatorname{det}(\mathbf{\Omega}) -\frac{1}{2} \mathbf{z}_i^{\top} \mathbf{\Omega} \mathbf{z}_i\right\}\right] \\
& = -\frac{p}{2} \log(2\pi) +\frac{1}{2}\log\operatorname{det}(\mathbf{\Omega}) -\frac{1}{2}  \mathbb{E}_{\mathbb{{\mathbb{Q}}}}\left[\mathbf{z}_i^{\top} \mathbf{\Omega} \mathbf{z}_i\right] \\
& = -\frac{p}{2} \log(2\pi) +\frac{1}{2}\log\operatorname{det}(\mathbf{\Omega}) -\frac{1}{2} \left[\tilde{\mathbf{z}}_i^{\top} \mathbf{\Omega} \tilde{\mathbf{z}}_i + \operatorname{tr}(\mathbf{\Omega} \tilde{\mathbf{\Omega}}^{-1})\right] \\
\end{aligned}
$$

$$
\begin{aligned}
\operatorname{KL}\left[\mathbb{Q}\left(\mathbf{z}_{i}\right) \| \mathbb{P}\left(\mathbf{z}_{i}\right)\right] & =\mathbb{E}_{\mathbb{{\mathbb{Q}}}}\left[\log {\mathbb{Q}}\left(\mathbf{z}_i \mid \mathbf{\Omega}\right)\right] - \mathbb{E}_{\mathbb{{\mathbb{Q}}}}\left[\log {\mathbb{P}}\left(\mathbf{z}_i \mid \mathbf{\Omega}\right)\right] \\
& =   \left\{ -\frac{p}{2}\log(2\pi) + \frac{1}{2}\log \operatorname{det}(\tilde{\mathbf{\Omega}}) -\frac{p}{2} \right\}\\ 
& \qquad - \left\{ -\frac{p}{2} \log(2\pi) +\frac{1}{2}\log\operatorname{det}(\mathbf{\Omega})  -\frac{1}{2} \left[\tilde{\mathbf{z}}_i^{\top} \mathbf{\Omega} \tilde{\mathbf{z}}_i + \operatorname{tr}(\mathbf{\Omega} \tilde{\mathbf{\Omega}}^{-1})\right] \right\}\\
&= -\frac{p}{2} + \frac{1}{2} \log \operatorname{det}(\mathbf{\Omega}^{-1} \tilde{\mathbf{\Omega}})  +\frac{1}{2} \left[\tilde{\mathbf{z}}_i^{\top} \mathbf{\Omega} \tilde{\mathbf{z}}_i + \operatorname{tr}(\mathbf{\Omega} \tilde{\mathbf{\Omega}}^{-1})\right] \\
&= -\frac{p}{2} + \frac{1}{2} \log \operatorname{det}(\mathbf{\Omega}^{-1} \tilde{\mathbf{\Omega}})   +\frac{1}{2} \tilde{\mathbf{z}}_i^{\top} \mathbf{\Omega} \tilde{\mathbf{z}}_i +\frac{1}{2} \operatorname{tr}(\mathbf{\Omega} \tilde{\mathbf{\Omega}}^{-1}) \\
\end{aligned}
$$
\clearpage
\noindent
Here follows the derivation of $\operatorname{KL}\left[\mathbb{Q}(\delta_1) \| \mathbb{P}(\delta_1)\right]$:

$$
\begin{aligned}
\mathbb{E}_{\mathbb{{\mathbb{Q}}}}\left[\log {\mathbb{Q}}\left(\delta_1\right)\right]&=\mathbb{E}_{\mathbb{{\mathbb{Q}}}}\left[\log \left\{\frac{\tilde{b}_1^{\tilde{a}_1}}{\Gamma\left(\tilde{a}_1\right)} \delta_1^{\tilde{a}_1-1} \exp \left(-\tilde{b}_1 \delta_1\right)\right\}\right] \\
&=\tilde{a}_1 \log \tilde{b}_1-\log \Gamma\left(\tilde{a}_1\right)+\left(\tilde{a}_1-1\right) \mathbb{E}_{\mathbb{{\mathbb{Q}}}}\left[\log \left(\delta_1\right)\right]-\tilde{b}_1\mathbb{E}_{\mathbb{{\mathbb{Q}}}}\left[\delta_1\right] \\
&=\tilde{a}_1 \log \tilde{b}_1-\log \Gamma\left(\tilde{a}_1\right)+\left(\tilde{a}_1-1\right) \left[\psi(\tilde{a}_1)+\log(\tilde{b}_1)\right]-\tilde{b}_1\left(\frac{\tilde{a}_1}{\tilde{b}_1}\right) \\
&=\tilde{a}_1 \log \tilde{b}_1-\log \Gamma\left(\tilde{a}_1\right)+\left(\tilde{a}_1-1\right) \left[\psi(\tilde{a}_1)+\log(\tilde{b}_1)\right]-\tilde{a}_1 \\
\mathbb{E}_{\mathbb{{\mathbb{Q}}}}\left[\log {\mathbb{P}}\left(\delta_1\right)\right] &=\mathbb{E}_{\mathbb{{\mathbb{Q}}}}\left[\log \left\{\frac{b_1^{a_1}}{\Gamma\left(a_1\right)} \delta_1^{a_1-1} \exp \left(-b_1 \delta_1\right)\right\}\right] \\
& =\mathbb{E}_{\mathbb{{\mathbb{Q}}}}\left[a_1 \log b_1-\log \Gamma\left(a_1\right)+\left(a_1-1\right) \log \left(\delta_1\right)-b_1\delta_1\right] \\
& =a_1 \log b_1-\log \Gamma\left(a_1\right)+\left(a_1-1\right) \mathbb{E}_{\mathbb{{\mathbb{Q}}}}\left[\log (\delta_1)\right]-b_1\mathbb{E}_{\mathbb{{\mathbb{Q}}}}\left[\delta_1\right] \\
& =a_1 \log b_1-\log \Gamma\left(a_1\right)+\left(a_1-1\right) \left[\psi(\tilde{a}_1)+\log(\tilde{b}_1)\right]-b_1\left(\frac{\tilde{a}_1}{\tilde{b}_1}\right) \\
\end{aligned}
$$
$$
\begin{aligned}
\operatorname{KL}\left[\mathbb{Q}(\delta_1) \| \mathbb{P}(\delta_1)\right] & =\mathbb{E}_{\mathbb{{\mathbb{Q}}}} \left[\log{\mathbb{Q}}\left(\delta_1\right)\right] - \mathbb{E}_{\mathbb{{\mathbb{Q}}}}\left[\log {\mathbb{P}}\left(\delta_1\right)\right]  \\
&=\left\{\tilde{a}_1 \log \tilde{b}_1-\log \Gamma\left(\tilde{a}_1\right)+\left(\tilde{a}_1-1\right) \left[\psi(\tilde{a}_1)+\log(\tilde{b}_1)\right]-\tilde{a}_1  \right\} \\ 
& \quad - \left\{ a_1 \log b_1-\log \Gamma\left(a_1\right)+\left(a_1-1\right) \left[\psi(\tilde{a}_1)+\log(\tilde{b}_1)\right] -b_1\left(\frac{\tilde{a}_1}{\tilde{b}_1}\right) \right\} \\
&= \tilde{a}_1  \left[\psi(\tilde{a}_1)+\frac{b_1}{\tilde{b}_1}+1\right] - a_1  \left[\psi(\tilde{a}_1)+\log(\frac{b_1}{\tilde{b}_1})\right] - \log\frac{\Gamma(\tilde{a}_1)}{\Gamma(a_1)}
\end{aligned}
$$

\clearpage
\noindent
Here follows the derivation of $\mathbb{E}_{\mathbb{Q}(\mathbf{Z}, \alpha \mid \mathbf{Y})}\left[\log (\mathbb{P}(\mathbf{Y} \mid \mathbf{Z}, \alpha)\right] $:

$$
\begin{aligned}
\mathbb{E}_{\mathbb{Q}(\mathbf{Z}, \alpha \mid \mathbf{Y})}[\log (\mathbb{P}(\mathbf{Y} \mid \mathbf{Z}, \alpha))]&= \mathbb{E}_{\mathbb{Q}(\mathbf{Z}, \alpha \mid \mathbf{Y})}\left[ \log \left\{ \prod_{i \neq j}^{n} \frac{\exp\left[y_{i,j}(\alpha-\left\|\mathbf{z}_{i}-\mathbf{z}_{j}\right\|^{2})\right]}{1+\exp(\alpha-\left\|\mathbf{z}_{i}-\mathbf{z}_{j}\right\|^{2})} \right\}\right] \\
&=\mathbb{E}_{\mathbb{Q}(\mathbf{Z}, \alpha \mid \mathbf{Y})}\left[\sum_{i \neq j}^{n} \left\{y_{i,j}\alpha-y_{i,j}\left\|\mathbf{z}_{i}-\mathbf{z}_{j}\right\|^{2} \right. \right. \\
& \quad \left. \left. -\log \left[1+\exp \left(\alpha-\left\|\mathbf{z}_{i}-\mathbf{z}_{j}\right\|^{2}\right)\right]\right\}\right] \\
&=\sum_{i \neq j}^{n} y_{i,j} \mathbb{E}_{\mathbb{Q}(\mathbf{Z}, \alpha \mid \mathbf{Y})}\left[\alpha-\left\|\mathbf{z}_{i}-\mathbf{z}_{j}\right\|^{2}\right] \\
& \quad -\mathbb{E}_{\mathbb{Q}(\mathbf{Z}, \alpha \mid \mathbf{Y})}\left[\log \left(1+\exp \left(\alpha-\left\|\mathbf{z}_{i}-\mathbf{z}_{j}\right\|^{2}\right)\right)\right] \\
&=\sum_{i \neq j}^{n} y_{i,j} \mathbb{E}_{\mathbb{Q}}\left[\alpha\right]-\mathbb{E}_{\mathbb{Q}}\left[\left\|\mathbf{z}_{i}-\mathbf{z}_{j}\right\|^{2}\right] \\
& \quad -\mathbb{E}_{\mathbb{Q}(\mathbf{Z}, \alpha \mid \mathbf{Y})}\left[\log \left(1+\exp \left(\alpha-\left\|\mathbf{z}_{i}-\mathbf{z}_{j}\right\|^{2}\right)\right)\right] \\
& =\sum_{i \neq j}^{n} y_{i,j}\left(\tilde{\mu}_{\alpha}-2 \operatorname{tr}(\tilde{\mathbf{\Omega}}^{-1})-\left\|\tilde{\mathbf{z}}_{i}-\tilde{\mathbf{z}}_{j}\right\|^{2}\right) \\
& \quad -\mathbb{E}_{\mathbb{Q}(\mathbf{Z}, \alpha \mid \mathbf{Y})}\left[\log \left(1+\exp \left(\alpha-\left\|\mathbf{z}_{i}-\mathbf{z}_{j}\right\|^{2}\right)\right)\right] \\
& \text{since log is concave, by Jensen inequality:} \\
& \leq \sum_{i \neq j}^{n} y_{i,j}\left(\tilde{\mu}_{\alpha}-2 \operatorname{tr}(\tilde{\mathbf{\Omega}}^{-1})-\left\|\tilde{\mathbf{z}}_{i}-\tilde{\mathbf{z}}_{j}\right\|^{2}\right) \\
& \quad -\log \left(1+\mathbb{E}_{\mathbb{Q}(\mathbf{Z}, \alpha \mid \mathbf{Y})}\left[\exp \left(\alpha-\left\|\mathbf{z}_{i}-\mathbf{z}_{j}\right\|^{2}\right)\right]\right) \\
& =\sum_{i \neq j}^{n} y_{i,j}\left(\tilde{\mu}_{\alpha}-2 \operatorname{tr}(\tilde{\mathbf{\Omega}}^{-1})-\left\|\tilde{\mathbf{z}}_{i}-\tilde{\mathbf{z}}_{j}\right\|^{2}\right) \\
& \quad -\log \left\{1+\frac{\exp \left(\tilde{\mu}_{\alpha}+\frac{1}{2} \tilde{\sigma}_{\alpha}^{2}\right)}{\operatorname{det}(\mathbf{I}+4 \tilde{\mathbf{\Omega}}^{-1})^{\frac{1}{2}}} \right. \\
& \qquad \left. \exp \left[-\left(\tilde{\mathbf{z}}_{i}-\tilde{\mathbf{z}}_{j}\right)^{\top}(\mathbf{I}+4 \tilde{\mathbf{\Omega}}^{-1})^{-1}\left(\tilde{\mathbf{z}}_{i}-\tilde{\mathbf{z}}_{j}\right)\right]\right\} .
\end{aligned}
$$

$$
\begin{aligned}
    \mathbb{E}_{\mathbb{Q}(\mathbf{Z}, \alpha \mid \mathbf{Y})}\left[\exp \left(\alpha-\left\|\mathbf{z}_{i}-\mathbf{z}_{j}\right\|^{2}\right)\right] = & \mathbb{E}_{\mathbb{Q}} \left[ \exp(\alpha)\right] \mathbb{E}_{\mathbb{Q}} \left[ \exp(-\| \tilde{\mathbf{z}}_i-\tilde{\mathbf{z}}_j \|)^2\right] \\
    = & \left[ \exp \left(\tilde{\mu}_{\alpha}+\frac{1}{2} \tilde{\sigma}_{\alpha}^{2}\right) \right] \mathbb{E}_{\mathbb{Q}} \left[ \exp(-\| \tilde{\mathbf{z}}_i-\tilde{\mathbf{z}}_j \|)^2\right] \\
    = & \left[ \exp \left(\tilde{\mu}_{\alpha}+\frac{1}{2} \tilde{\sigma}_{\alpha}^{2}\right) \right] \left\{\operatorname{det}\left( 4 \tilde{\mathbf{\Omega}}^{-1} +\mathbf{I} \right)^{-1/2} \right. \\
    & \left. \exp \left[ -\left(\tilde{\mathbf{z}}_i-\tilde{\mathbf{z}}_j\right)^\top   \left( 4 \tilde{\mathbf{\Omega}}^{-1} + \mathbf{I} \right)^{-1} \left(\tilde{\mathbf{z}}_i-\tilde{\mathbf{z}}_j\right) \right] \right\} \\
\end{aligned}
$$

$$
\begin{aligned}
    \mathbb{Q}\left(\mathbf{z}_i\right) \sim \operatorname{MVN}&\left(\tilde{\mathbf{z}}_i, \tilde{\mathbf{\Omega}}^{-1}\right)  \qquad \mathbb{Q}\left(\mathbf{z}_j\right) \sim \operatorname{MVN}\left(\tilde{\mathbf{z}}_j, \tilde{\mathbf{\Omega}}^{-1}\right) \\
    \therefore \mathbb{Q}\left(\mathbf{z}_i-\mathbf{z}_j\right) &\sim \operatorname{MVN}\left(\tilde{\mathbf{z}}_i-\tilde{\mathbf{z}}_j, 2 \tilde{\mathbf{\Omega}}^{-1}\right) \\
\text { let } \mathbf{D}_{i,j}=\mathbf{z}_i-\mathbf{z}_j, \tilde{\mathbf{S}}&=2 \tilde{\mathbf{\Omega}}^{-1} \qquad \therefore  \mathbb{Q}\left(\mathbf{D}_{i,j}\right) \sim \operatorname{MVN}\left(\tilde{\mathbf{D}}_{i,j}, \tilde{\mathbf{S}}\right) \\
\end{aligned}$$

$$
\begin{aligned}
\mathbb{E}_{\mathbb{Q}}\left[\exp \left(-\mathbf{D}_{i,j}^\top \mathbf{D}_{i,j}\right)\right]=&\int_{-\infty}^{\infty} \exp \left(-\mathbf{D}_{i,j}^\top \mathbf{D}_{i,j}\right) \operatorname{det}(2 \pi \tilde{\mathbf{S}})^{-1 / 2} \\
& \exp \left[-\frac{1}{2}\left(\mathbf{D}_{i,j}-\tilde{\mathbf{D}}_{i,j}\right)^\top \tilde{\mathbf{S}}^{-1} \left(\mathbf{D}_{i,j}-\tilde{\mathbf{D}}_{i,j}\right) \right] d \mathbf{D}_{i,j} \\
=& \operatorname{det}(2 \pi \tilde{\mathbf{S}})^{-1 / 2} \int_{-\infty}^{\infty} \exp \left[-\mathbf{D}_{i,j}^\top \mathbf{D}_{i,j}-\frac{1}{2} \mathbf{D}_{i,j}^\top \tilde{\mathbf{S}}^{-1}\mathbf{D}_{i,j} \right. \\
& \left. +\mathbf{D}_{i,j}^\top \tilde{\mathbf{S}}^{-1}\tilde{\mathbf{D}}_{i,j} -\frac{1}{2} \tilde{\mathbf{D}}_{i,j}^\top \tilde{\mathbf{S}}^{-1}\tilde{\mathbf{D}}_{i,j} \right] d \mathbf{D}_{i,j} \\
=& \operatorname{det}(2 \pi \tilde{\mathbf{S}})^{-1 / 2} \exp \left[-\frac{1}{2} \tilde{\mathbf{D}}_{i,j}^\top \tilde{\mathbf{S}}^{-1}\tilde{\mathbf{D}}_{i,j} \right] \\
&\int_{-\infty}^{\infty} \exp \left[-\mathbf{D}_{i,j}^\top \mathbf{D}_{i,j} -\frac{1}{2} \mathbf{D}_{i,j}^\top \tilde{\mathbf{S}}^{-1}\mathbf{D}_{i,j}+\mathbf{D}_{i,j}^\top \tilde{\mathbf{S}}^{-1}\tilde{\mathbf{D}}_{i,j}  \right] d \mathbf{D}_{i,j} \\
=& \operatorname{det}(2 \pi \tilde{\mathbf{S}})^{-1 / 2} \exp \left[-\frac{1}{2} \tilde{\mathbf{D}}_{i,j}^\top \tilde{\mathbf{S}}^{-1}\tilde{\mathbf{D}}_{i,j} \right] \\
&\int_{-\infty}^{\infty} \exp \left\{-\frac{1}{2}\mathbf{D}_{i,j}^\top \left[ \left(\frac{1}{2} \mathbf{I}\right)^{-1} + \tilde{\mathbf{S}}^{-1} \right]\mathbf{D}_{i,j}+\mathbf{D}_{i,j}^\top \tilde{\mathbf{S}}^{-1}\tilde{\mathbf{D}}_{i,j}  \right\} d \mathbf{D}_{i,j} \\
\end{aligned}
$$

$$
\begin{aligned}
\mathbb{E}_{\mathbb{Q}} &\left[\exp \left(-\mathbf{D}_{i,j}^\top \mathbf{D}_{i,j}\right)\right] \\
=& \operatorname{det}(2 \pi \tilde{\mathbf{S}})^{-1 / 2} \exp \left[-\frac{1}{2} \tilde{\mathbf{D}}_{i,j}^\top \tilde{\mathbf{S}}^{-1}\tilde{\mathbf{D}}_{i,j} \right]  \operatorname{det}\left\{ 2\pi \left[ \left( \frac{1}{2}\mathbf{I}\right)^{-1} + \tilde{\mathbf{S}}^{-1} \right]^{-1}\right\}^{1/2} \\
& \exp \left\{\frac{1}{2}\tilde{\mathbf{D}}_{i,j}^\top \tilde{\mathbf{S}}^{-1}\left[ \left(\frac{1}{2} \mathbf{I}\right)^{-1} + \tilde{\mathbf{S}}^{-1} \right]^{-1}\tilde{\mathbf{S}}^{-1}\tilde{\mathbf{D}}_{i,j}\right\} \\
=& \operatorname{det}(2 \pi \tilde{\mathbf{S}})^{-1 / 2} \operatorname{det}\left\{ 2\pi \left[ \left( \frac{1}{2}\mathbf{I}\right)^{-1} + \tilde{\mathbf{S}}^{-1} \right]^{-1}\right\}^{1/2} \\
&\exp \left[-\frac{1}{2} \tilde{\mathbf{D}}_{i,j}^\top \tilde{\mathbf{S}}^{-1}\tilde{\mathbf{D}}_{i,j} \right]  \exp \left\{\frac{1}{2}\tilde{\mathbf{D}}_{i,j}^\top \tilde{\mathbf{S}}^{-1}\left[ \left(\frac{1}{2} \mathbf{I}\right)^{-1} + \tilde{\mathbf{S}}^{-1} \right]^{-1}\tilde{\mathbf{S}}^{-1}\tilde{\mathbf{D}}_{i,j}\right\} \\
=& \operatorname{det}\left\{ \tilde{\mathbf{S}}\left[ 2\mathbf{I} + \tilde{\mathbf{S}}^{-1} \right]\right\}^{-1/2}\exp \left\{-\frac{1}{2} \tilde{\mathbf{D}}_{i,j}^\top \tilde{\mathbf{S}}^{-1}\tilde{\mathbf{D}}_{i,j} \right. \\
& \left. +\frac{1}{2}\tilde{\mathbf{D}}_{i,j}^\top \left[ 2\tilde{\mathbf{S}}\tilde{\mathbf{S}} + \tilde{\mathbf{S}}\tilde{\mathbf{S}}^{-1}\tilde{\mathbf{S}} \right]^{-1}\tilde{\mathbf{D}}_{i,j}\right\} \\
=& \operatorname{det} \left( 2\tilde{\mathbf{S}} + \mathbf{I}  \right)^{-1/2}\exp \left\{ \tilde{\mathbf{D}}_{i,j}^\top \left(\frac{1}{2}\tilde{\mathbf{S}}^{-1}\right) \left[-\mathbf{I} + \left( 2\tilde{\mathbf{S}} + \mathbf{I} \right)^{-1} \right]\tilde{\mathbf{D}}_{i,j} \right\} \\
=& \operatorname{det} \left( 2\tilde{\mathbf{S}} + \mathbf{I}  \right)^{-1/2}\exp \left\{ \tilde{\mathbf{D}}_{i,j}^\top \left(\frac{1}{2}\tilde{\mathbf{S}}^{-1}\right) \left[\left(-\mathbf{I} - 2\tilde{\mathbf{S}} + \mathbf{I} \right) \left( 2\tilde{\mathbf{S}} + \mathbf{I} \right)^{-1} \right]\tilde{\mathbf{D}}_{i,j} \right\} \\
=& \operatorname{det} \left( 2\tilde{\mathbf{S}} + \mathbf{I}  \right)^{-1/2}\exp \left\{ -\tilde{\mathbf{D}}_{i,j}^\top   \left( 2\tilde{\mathbf{S}} + \mathbf{I} \right)^{-1} \tilde{\mathbf{D}}_{i,j} \right\} \\
\\ 
\therefore \mathbb{E}_{\mathbb{Q}}& \left[ \exp(-\| \tilde{\mathbf{z}}_i-\tilde{\mathbf{z}}_j \|)^2\right] \\
=& \operatorname{det}\left( 4 \tilde{\mathbf{\Omega}}^{-1} +\mathbf{I} \right)^{-1/2}\exp \left\{ -\left(\tilde{\mathbf{z}}_i-\tilde{\mathbf{z}}_j\right)^\top   \left( 4 \tilde{\mathbf{\Omega}}^{-1} + \mathbf{I} \right)^{-1} \left(\tilde{\mathbf{z}}_i-\tilde{\mathbf{z}}_j\right) \right\} \\
\end{aligned}
$$

\clearpage

To sum up, the KL divergence is:
$$
\begin{aligned}
\mathrm{KL}&[\mathbb{Q}(\alpha, \mathbf{Z}, \bm{\delta}) \| \mathbb{P}(\alpha, \mathbf{Z}, \bm{\delta} \mid \mathbf{Y})] \\
 = &  \operatorname{KL}\left[\mathbb{Q}(\alpha) \| \mathbb{P}(\alpha)\right]  + \sum_{i=1}^{n} \operatorname{KL}\left[\mathbb{Q}\left(\mathbf{z}_{i}\right) \| \mathbb{P}\left(\mathbf{z}_{i}\right)\right] +\operatorname{KL}\left[\mathbb{Q}(\delta_1) \| \mathbb{P}(\delta_1)\right]  \\
&  +\sum_{h=2}^p \operatorname{KL}\left[\mathbb{Q}\left(\delta_{h}\right) \| \mathbb{P}\left(\delta_{h}\right)\right] -\mathbb{E}_{\mathbb{Q}(\mathbf{Z}, \alpha \mid \mathbf{Y})}\left[\log (\mathbb{P}(\mathbf{Y} \mid \mathbf{Z}, \alpha)\right] +\log \mathbb{P}(\mathbf{Y}) \\
\leq &-\frac{1}{2} \log \left( \frac{\tilde{\sigma}_\alpha^2}{\sigma_\alpha^2 }\right)-\frac{1}{2} +\frac{\tilde{\sigma}_\alpha^2}{2\sigma_\alpha^2} +\frac{(\tilde{\mu}_\alpha-\mu_\alpha)^2}{2\sigma_\alpha^2}   \\
& -\frac{np}{2} + \frac{n}{2} \log \operatorname{det}(\mathbf{\Omega}^{-1} \tilde{\mathbf{\Omega}})   +\frac{1}{2} \sum_{i=1}^{n} \tilde{\mathbf{z}}_i^{\top} \mathbf{\Omega} \tilde{\mathbf{z}}_i +\frac{n}{2} \operatorname{tr}(\mathbf{\Omega} \tilde{\mathbf{\Omega}}^{-1}) \\
&+ \tilde{a}_1  \left[\psi(\tilde{a}_1)+\frac{b_1}{\tilde{b}_1}+1\right] - a_1  \left[\psi(\tilde{a}_1)+\log(\frac{b_1}{\tilde{b}_1})\right] - \log\frac{\Gamma(\tilde{a}_1)}{\Gamma(a_1)} \\
& +\sum_{h=2}^{p} \operatorname{KL}\left[\mathbb{Q}\left(\delta_{h}\right) \| \mathbb{P}\left(\delta_{h}\right)\right] \\
& -\sum_{i \neq j}^{n} y_{i,j}\left(\tilde{\mu}_{\alpha}-2 \operatorname{tr}(\tilde{\mathbf{\Omega}}^{-1})-\left\|\tilde{\mathbf{z}}_{i}-\tilde{\mathbf{z}}_{j}\right\|^{2}\right) \\
& \quad +\log \left\{1+\frac{\exp \left(\tilde{\mu}_{\alpha}+\frac{1}{2} \tilde{\sigma}_{\alpha}^{2}\right)}{\operatorname{det}(\mathbf{I}+4 \tilde{\mathbf{\Omega}}^{-1})^{\frac{1}{2}}} \right. \\
& \qquad \left. \exp \left[-\left(\tilde{\mathbf{z}}_{i}-\tilde{\mathbf{z}}_{j}\right)^{\top}(\mathbf{I}+4 \tilde{\mathbf{\Omega}}^{-1})^{-1}\left(\tilde{\mathbf{z}}_{i}-\tilde{\mathbf{z}}_{j}\right)\right]\right\} \\
&  +\log \mathbb{P}(\mathbf{Y}) \\
\end{aligned}
$$

\clearpage
\section{Summary for updating the variational parameters}
\label{app:vlspm-partial}

The variational parameters are updated in turns, while holding the others constant as follow:

\begin{enumerate}

    \item Optimize $\tilde{\mu}_{\alpha}$ via bracketing-and-bisection algorithm on the following expression:
\begin{equation}
\begin{aligned}
\frac{\partial \operatorname{KL}}{\partial \tilde{\mu}_{\alpha} } &= \frac{(\tilde{\mu}_\alpha-\mu_\alpha)}{\sigma_\alpha^2} -\sum_{i \neq j}^{n} y_{i,j} +\sum_{i \neq j}^{n} \left[\exp \left(-\tilde{\mu}_{\alpha} \right) A_{i,j}^{-1} + 1\right]^{-1} \\
\text{where } A_{i,j} &= \exp \left(\frac{1}{2} \tilde{\sigma}_\alpha^2\right) \operatorname{det}\left(\mathbf{I}+4 \tilde{\bm{\Omega}}^{-1}\right)^{-1 / 2} \exp \left[-\left(\tilde{\mathbf{z}}_i-\tilde{\mathbf{z}}_j\right)^{\top}\left(\mathbf{I}+4 \tilde{\bm{\Omega}}^{-1}\right)^{-1}\left(\tilde{\mathbf{z}}_i-\tilde{\mathbf{z}}_j\right)\right] \\
\end{aligned}
\end{equation}

\item Optimize $\tilde{\sigma}^2_{\alpha}$ via bracketing-and-bisection algorithm on the following expression:
\begin{equation}
\begin{aligned}
\frac{\partial \operatorname{KL}}{\partial \tilde{\sigma}_{\alpha}^2 } &= \frac{1}{2\sigma_{\alpha}^2} - \frac{1}{2\tilde{\sigma}_{\alpha}^2} + \sum_{i \neq j}^n \frac{1}{2} \left[ 1 + \exp(-\frac{1}{2}\tilde{\sigma}_{\alpha}^2)B_{i,j}^{-1}\right]^{-1} \\
\text{where } B_{i,j}&=\exp \left(\tilde{\mu}_\alpha\right) \operatorname{det}\left(\mathbf{I}+4 \tilde{\bm{\Omega}}^{-1}\right)^{-1 / 2} \exp \left[-\left(\tilde{\mathbf{z}}_i-\tilde{\mathbf{z}}_j\right)^{\top}\left(\mathbf{I}+4 \tilde{\bm{\Omega}}^{-1}\right)^{-1}\left(\tilde{\mathbf{z}}_i-\tilde{\mathbf{z}}_j\right)\right] \\
\end{aligned}
\end{equation}

\item Optimize $\tilde{\mathbf{z}}_i$ via conjugate gradient routine on the following expression:
\begin{equation}
\begin{aligned}
\frac{\partial \operatorname{KL}}{\partial \tilde{\mathbf{z}}_i} &= 2\tilde{\mathbf{z}}_i \left(\frac{1}{2} \bm{\Omega} + \sum_{i \neq j}^n y_{i,j} \right) - 2 \left(\sum_{i \neq j}^n y_{i,j}  \tilde{\mathbf{z}}_j \right) + 2 G(\tilde{\mathbf{z}}_i) \\
\text{where } G\left(\tilde{\mathbf{z}}_i\right)&=-2\left(\mathbf{I}+4 \tilde{\bm{\Omega}}^{-1}\right) \sum_{i \neq j}^n\left(\tilde{\mathbf{z}}_i-\tilde{\mathbf{z}}_j\right)\left\{1+\frac{\operatorname{det}\left(\mathbf{I}+\tilde{\bm{\Omega}}^{-1}\right)^{1 / 2}}{\exp \left(\tilde{\mu}_\alpha+\frac{1}{2} \tilde{\sigma}_\alpha^2\right)} \right. \\
& \quad \left. \exp \left[-\left(\tilde{\mathbf{z}}_i-\tilde{\mathbf{z}}_j\right)^{\top}\left(\mathbf{I}+4 \tilde{\bm{\Omega}}^{-1}\right)^{-1}\left(\tilde{\mathbf{z}}_i-\tilde{\mathbf{z}}_j\right)\right]\right\}^{-1}
\\
\end{aligned}
\end{equation}

\item Updates $\delta_1$ analytically with the following hyperparameters:
$$
\tilde{a}_1 = \frac{np}{2}+a_1
\qquad
\tilde{b}_1 = \quad b_1 + \frac{1}{2}\sum_{i=1}^{n}\sum_{\ell=1}^{p} \prod_{m=2}^{\ell} 
\mathbb{E}(\delta_m) \left[ \tilde{z}_{i\ell}^2 + \tilde{\omega}_{\ell}^{-1}\right] 
$$

\item Updates $\delta_h$ for $h=2, \ldots, p$ analytically with the following hyperparameters:
$$
\tilde{a}_2^h = \frac{n(p-h+1)}{2}+a_2
\qquad
\tilde{b}_2^h = \quad b_2 + \frac{1}{2}\sum_{i=1}^{n}\sum_{\ell=h}^{p} \prod_{m=1, m \neq h}^{\ell} 
\mathbb{E}(\delta_m) \left[ \tilde{z}_{i\ell}^2 + \tilde{\omega}_{\ell}^{-1}\right]
$$
\end{enumerate}


\section{Derivations of the optimal variational parameters} \label{app:vlspm-fullopt}
The full approximate KL divergence needed for the starting point of the partial derivatives is as follow:
$$
\begin{aligned}
\mathrm{KL}\left[\mathbb{Q}(\alpha, \mathbf{Z}, \bm{\delta}) \| \mathbb{P}(\alpha, \mathbf{Z}, \bm{\delta} \mid \mathbf{Y})\right]
\leq &-\frac{1}{2} \log \left( \frac{\tilde{\sigma}_\alpha^2}{\sigma_\alpha^2 }\right)-\frac{1}{2} +\frac{\tilde{\sigma}_\alpha^2}{2\sigma_\alpha^2} +\frac{(\tilde{\mu}_\alpha-\mu_\alpha)^2}{2\sigma_\alpha^2}   \\
& -\frac{np}{2} + \frac{n}{2} \log \operatorname{det}(\mathbf{\Omega}^{-1} \tilde{\mathbf{\Omega}})   +\frac{1}{2} \sum_{i=1}^{n} \tilde{\mathbf{z}}_i^{\top} \mathbf{\Omega} \tilde{\mathbf{z}}_i +\frac{n}{2} \operatorname{tr}(\mathbf{\Omega} \tilde{\mathbf{\Omega}}^{-1}) \\
&+ \tilde{a}_1  \left[\psi(\tilde{a}_1)+\frac{b_1}{\tilde{b}_1}+1\right] - a_1  \left[\psi(\tilde{a}_1)+\log(\frac{b_1}{\tilde{b}_1})\right] - \log\frac{\Gamma(\tilde{a}_1)}{\Gamma(a_1)} \\
& +\sum_{h=2}^{p} \operatorname{KL}\left[\mathbb{Q}\left(\delta_{h}\right) \| \mathbb{P}\left(\delta_{h}\right)\right] \\
& -\sum_{i \neq j}^{n} y_{i,j}\left(\tilde{\mu}_{\alpha}-2 \operatorname{tr}(\tilde{\mathbf{\Omega}}^{-1})-\left\|\tilde{\mathbf{z}}_{i}-\tilde{\mathbf{z}}_{j}\right\|^{2}\right) \\
& \quad +\log \left\{1+\frac{\exp \left(\tilde{\mu}_{\alpha}+\frac{1}{2} \tilde{\sigma}_{\alpha}^{2}\right)}{\operatorname{det}(\mathbf{I}+4 \tilde{\mathbf{\Omega}}^{-1})^{\frac{1}{2}}} \right. \\
& \qquad \left. \exp \left[-\left(\tilde{\mathbf{z}}_{i}-\tilde{\mathbf{z}}_{j}\right)^{\top}(\mathbf{I}+4 \tilde{\mathbf{\Omega}}^{-1})^{-1}\left(\tilde{\mathbf{z}}_{i}-\tilde{\mathbf{z}}_{j}\right)\right]\right\}   +\log \mathbb{P}(\mathbf{Y}) \\
\end{aligned}
$$

The partial derivatives for the variational parameter $\tilde{\mu}_\alpha$:
$$
\begin{aligned}
\mathrm{KL}_{\tilde{\mu}_\alpha}\leq &\frac{(\tilde{\mu}_\alpha-\mu_\alpha)^2}{2\sigma_\alpha^2} -\sum_{i \neq j}^{n} y_{i,j}\tilde{\mu}_{\alpha} + \sum_{i \neq j}^{n}\log \left[1+\exp \left(\tilde{\mu}_{\alpha} \right) A_{i,j}\right] \\
\text{where } A_{i,j} =& \frac{\exp \left(\frac{1}{2} \tilde{\sigma}_{\alpha}^{2}\right)}{\operatorname{det}(\mathbf{I}+4 \tilde{\mathbf{\Omega}}^{-1})^{\frac{1}{2}}} \exp \left[-\left(\tilde{\mathbf{z}}_{i}-\tilde{\mathbf{z}}_{j}\right)^{\top}(\mathbf{I}+4 \tilde{\mathbf{\Omega}}^{-1})^{-1}\left(\tilde{\mathbf{z}}_{i}-\tilde{\mathbf{z}}_{j}\right)\right] \\
\frac{\partial\mathrm{KL}_{\tilde{\mu}_\alpha}}{\partial \tilde{\mu}_{\alpha}}= &\frac{2(\tilde{\mu}_\alpha-\mu_\alpha)}{2\sigma_\alpha^2} -\sum_{i \neq j}^{n} y_{i,j} +\sum_{i \neq j}^{n} \left[ \frac{\exp \left(\tilde{\mu}_{\alpha} \right) A_{i,j}}{1+\exp \left(\tilde{\mu}_{\alpha} \right) A_{i,j}}\right] \\
= &\frac{(\tilde{\mu}_\alpha-\mu_\alpha)}{\sigma_\alpha^2} -\sum_{i \neq j}^{n} y_{i,j} +\sum_{i \neq j}^{n} \left[\exp \left(-\tilde{\mu}_{\alpha} \right) A_{i,j}^{-1} + 1\right]^{-1} \\
\end{aligned}
$$

The partial derivatives for the variational parameter $\tilde{\sigma}_\alpha^2$:
$$
\begin{aligned}
\mathrm{KL}_{\tilde{\sigma}_\alpha^2}\leq &-\frac{1}{2} \log \left( \frac{\tilde{\sigma}_\alpha^2}{\sigma_\alpha^2 }\right) +\frac{\tilde{\sigma}_\alpha^2}{2\sigma_\alpha^2} +\sum_{i \neq j}^{n} \log \left[ 1+\exp \left(\frac{1}{2} \tilde{\sigma}_{\alpha}^{2}\right)B_{i,j} \right] \\
\text{where } B_{i,j} = & \frac{\exp \left(\tilde{\mu}_{\alpha}\right)}{\operatorname{det}(\mathbf{I}+4 \tilde{\mathbf{\Omega}}^{-1})^{\frac{1}{2}}}  \exp \left[-\left(\tilde{\mathbf{z}}_{i}-\tilde{\mathbf{z}}_{j}\right)^{\top}(\mathbf{I}+4 \tilde{\mathbf{\Omega}}^{-1})^{-1}\left(\tilde{\mathbf{z}}_{i}-\tilde{\mathbf{z}}_{j}\right)\right] \\
\frac{\partial\mathrm{KL}_{\tilde{\sigma}_\alpha^2}}{\partial \tilde{\sigma}_{\alpha}^2}= &-\frac{1}{2\tilde{\sigma}_\alpha^2}  +\frac{1}{2\sigma_\alpha^2} +\sum_{i \neq j}^{n} \log \left[\exp \left(-\frac{1}{2} \tilde{\sigma}_{\alpha}^{2}\right)B_{i,j}^{-1} + 1 \right]^{-1} \\
\end{aligned}
$$
The partial derivatives for the variational parameter $\tilde{\mathbf{z}}_i$:
$$
\begin{aligned}
\mathrm{KL}_{\tilde{\mathbf{z}}_i}\leq & \frac{1}{2} \sum_{i=1}^{n} \tilde{\mathbf{z}}_i^{\top} \mathbf{\Omega} \tilde{\mathbf{z}}_i  -\sum_{i \neq j}^{n} y_{i,j}\left(-\left\|\tilde{\mathbf{z}}_{i}-\tilde{\mathbf{z}}_{j}\right\|^{2}\right) + G\left(\tilde{\mathbf{z}}_i\right) \\
= & \frac{1}{2} \sum_{i=1}^{n} \tilde{\mathbf{z}}_i^{\top} \mathbf{\Omega} \tilde{\mathbf{z}}_i  +\sum_{i \neq j}^{n} y_{i,j}\left(\tilde{\mathbf{z}}_{i}^{\top} \tilde{\mathbf{z}}_{i} -2\tilde{\mathbf{z}}_{i}^{\top}\tilde{\mathbf{z}}_{j}+ \tilde{\mathbf{z}}_{j}^{\top}\tilde{\mathbf{z}}_{j}\right) +G\left(\tilde{\mathbf{z}}_i\right) \\
\text{where } G\left(\tilde{\mathbf{z}}_i\right) =& \log \left\{1+\frac{\exp \left(\tilde{\mu}_{\alpha}+\frac{1}{2} \tilde{\sigma}_{\alpha}^{2}\right)}{\operatorname{det}(\mathbf{I}+4 \tilde{\mathbf{\Omega}}^{-1})^{\frac{1}{2}}}  \exp \left[-\left(\tilde{\mathbf{z}}_{i}-\tilde{\mathbf{z}}_{j}\right)^{\top}(\mathbf{I}+4 \tilde{\mathbf{\Omega}}^{-1})^{-1}\left(\tilde{\mathbf{z}}_{i}-\tilde{\mathbf{z}}_{j}\right)\right]\right\} \\
\text{note that if } F(x) =& \log \left[ 1 + a\exp(bX) \right], \text{then: } \\
F'(x) =& \left[ 1 + a\exp(bX) \right]^{-1}\left[a\exp(bX)\right]\left[b\right] = \left[ 1 + a^{-1}\exp(-bX) \right]^{-1}\left[b\right] \\
\therefore G'\left(\tilde{\mathbf{z}}_i\right)=&\left\{1+\frac{\operatorname{det}\left(\mathbf{I}+\tilde{\bm{\Omega}}^{-1}\right)^{1 / 2}}{\exp \left(\tilde{\mu}_\alpha+\frac{1}{2} \tilde{\sigma}_\alpha^2\right)}  \exp \left[-\left(\tilde{\mathbf{z}}_i-\tilde{\mathbf{z}}_j\right)^{\top}\left(\mathbf{I}+4 \tilde{\bm{\Omega}}^{-1}\right)^{-1}\left(\tilde{\mathbf{z}}_i-\tilde{\mathbf{z}}_j\right)\right]\right\}^{-1} \\
&\left\{-2\left(\mathbf{I}+4 \tilde{\bm{\Omega}}^{-1}\right)^{-1} \sum_{i \neq j}^n\left(\tilde{\mathbf{z}}_i-\tilde{\mathbf{z}}_j\right) \right\}
\\
\frac{\partial\mathrm{KL}_{\tilde{\mathbf{z}}_i}}{\partial \tilde{\mathbf{z}}_{i}}= & 2\tilde{\mathbf{z}}_i \left(\frac{1}{2} \bm{\Omega} + \sum_{i \neq j}^n y_{i,j} \right) - 2 \left(\sum_{i \neq j}^n y_{i,j}  \tilde{\mathbf{z}}_j \right) + G'(\tilde{\mathbf{z}}_i)\\
\end{aligned}
$$
Identifying the gamma density of $\delta_1$ to update for the variational parameters $\tilde{a}_1$ and $\tilde{b}_1$
$$
\begin{aligned}
\mathbb{P}(\delta_1 |-) \propto &  \delta_{1}^{\frac{np}{2}+a_1-1}
\exp\left[- \left(b_1 + \frac{1}{2}\sum_{i=1}^{n}\sum_{\ell=1}^{p} \prod_{m=2}^{\ell} \delta_{m} z_{i\ell}^2\right) \delta_1 \right] \\
\log \mathbb{P}(\delta_1 |-) \propto &  \left(\frac{np}{2}+a_1-1 \right) \log \left( \delta_{1} \right)
- \left(b_1 + \frac{1}{2}\sum_{i=1}^{n}\sum_{\ell=1}^{p} \prod_{m=2}^{\ell} \delta_{m} z_{i\ell}^2\right) \delta_1  \\
\mathbb{E}_{\mathbb{Q}} \left[ \log \mathbb{P}(\delta_1 |-) \right]  \propto & \left(\frac{np}{2}+a_1-1 \right) \log \left( \delta_{1} \right)
- \left(b_1 + \frac{1}{2}\sum_{i=1}^{n}\sum_{\ell=1}^{p} \prod_{m=2}^{\ell} \mathbb{E}_{\mathbb{Q}} \left[\delta_{m} \right] \mathbb{E}_{\mathbb{Q}} \left[z_{i\ell}^2\right]\right) \delta_1  \\
=  & \left(\frac{np}{2}+a_1-1 \right) \log \left( \delta_{1} \right)
- \left(b_1 + \frac{1}{2}\sum_{i=1}^{n}\sum_{\ell=1}^{p} \prod_{m=2}^{\ell}  \left[ \frac{\tilde{a}_2^m}{\tilde{b}_2^m} \right] \left[ \tilde{z}_{i\ell}^2 + \tilde{\omega}_{\ell}^{-1}\right]\right) \delta_1  \\
\delta_1 &\sim \text{Gam}\left(\frac{np}{2}+a_1, 
\quad b_1 + \frac{1}{2}\sum_{i=1}^{n}\sum_{\ell=1}^{p} \prod_{m=2}^{\ell} \frac{\tilde{a}_2^m}{\tilde{b}_2^m} \left[ \tilde{z}_{i\ell}^2 + \tilde{\omega}_{\ell}^{-1}\right] \right) \\
\therefore \tilde{a}_1 &= \frac{np}{2}+a_1,  \quad \tilde{b}_1 = b_1 + \frac{1}{2}\sum_{i=1}^{n}\sum_{\ell=1}^{p} \prod_{m=2}^{\ell} \frac{\tilde{a}_2^m}{\tilde{b}_2^m} \left[ \tilde{z}_{i\ell}^2 + \tilde{\omega}_{\ell}^{-1}\right]
\end{aligned}
$$
Identifying the truncated gamma density of $\delta_h$ to update for the variational parameters $\tilde{a}_1$ and $\tilde{b}_1$
$$
\begin{aligned}
\mathbb{P}(\delta_h |-)  \propto & \delta_{h}^{\frac{n(p-h+1)}{2}+a_2-1}    \exp\left[- \left(b_2+\frac{1}{2}\sum_{i=1}^{n}\sum_{\ell=h}^{p} \prod_{m=1, m \neq h}^{\ell} \delta_m  z_{i\ell}^2 \right)\delta_h\right] \\ 
\log \mathbb{P}(\delta_h |-)  \propto & \left( \frac{n(p-h+1)}{2}+a_2-1\right) \log \left(\delta_{h} \right)   - \left(b_2+\frac{1}{2}\sum_{i=1}^{n}\sum_{\ell=h}^{p} \prod_{m=1, m \neq h}^{\ell} \delta_m  z_{i\ell}^2 \right)\delta_h \\ 
\mathbb{E}_{\mathbb{Q}} \left[ \log \mathbb{P}(\delta_h |-) \right] \propto & \left( \frac{n(p-h+1)}{2}+a_2-1\right) \log \left(\delta_{h} \right)  \\
& - \left(b_2+\frac{1}{2}\sum_{i=1}^{n}\sum_{\ell=h}^{p} \prod_{m=1, m \neq h}^{\ell} \mathbb{E}_{\mathbb{Q}} \left[ \delta_m \right]  \mathbb{E}_{\mathbb{Q}} \left[z_{i\ell}^2 \right]\right)\delta_h \\ 
= & \left( \frac{n(p-h+1)}{2}+a_2-1\right) \log \left(\delta_{h} \right)  \\
&- \left(b_2+\frac{1}{2}\sum_{i=1}^{n}\sum_{\ell=h}^{p} \left[\frac{\tilde{a}_1}{\tilde{b}_1} \right] \prod_{m=2, m \neq h}^{\ell} \left[ \frac{\tilde{a}_2^m}{\tilde{b}_2^m} \right]  \left[ \tilde{z}_{i\ell}^2 + \tilde{\omega}_{\ell}^{-1}\right]\right)\delta_h \\ 
\end{aligned}
$$
$$
\begin{aligned}
&\text{since $\delta_h$ is bounded between $[1, \infty)$,}  \\
\delta_h &\sim \text{Gam}^{\top}\left(\frac{n(p-h+1)}{2}+a_2, 
b_2+\frac{1}{2}\sum_{i=1}^{n}\sum_{\ell=h}^{p} \left[\frac{\tilde{a}_1}{\tilde{b}_1} \right] \prod_{m=2, m \neq h}^{\ell} \left[ \frac{\tilde{a}_2^m}{\tilde{b}_2^m} \right]  \left[ \tilde{z}_{i\ell}^2 + \tilde{\omega}_{\ell}^{-1}\right] \right) \\
\therefore \tilde{a}_2^{(h)} &= \frac{n(p-h+1)}{2}+a_2,  \quad \tilde{b}_2^{(h)} = b_2+\frac{1}{2}\sum_{i=1}^{n}\sum_{\ell=h}^{p} \left[\frac{\tilde{a}_1}{\tilde{b}_1} \right] \prod_{m=2, m \neq h}^{\ell} \left[ \frac{\tilde{a}_2^m}{\tilde{b}_2^m} \right]  \left[ \tilde{z}_{i\ell}^2 + \tilde{\omega}_{\ell}^{-1}\right]
\end{aligned}
$$

\clearpage
\section{Latent positions of the worm network} \label{app:vlspm-pos}
\begin{figure}[htb]
    \centering
    \includegraphics[width=\linewidth]{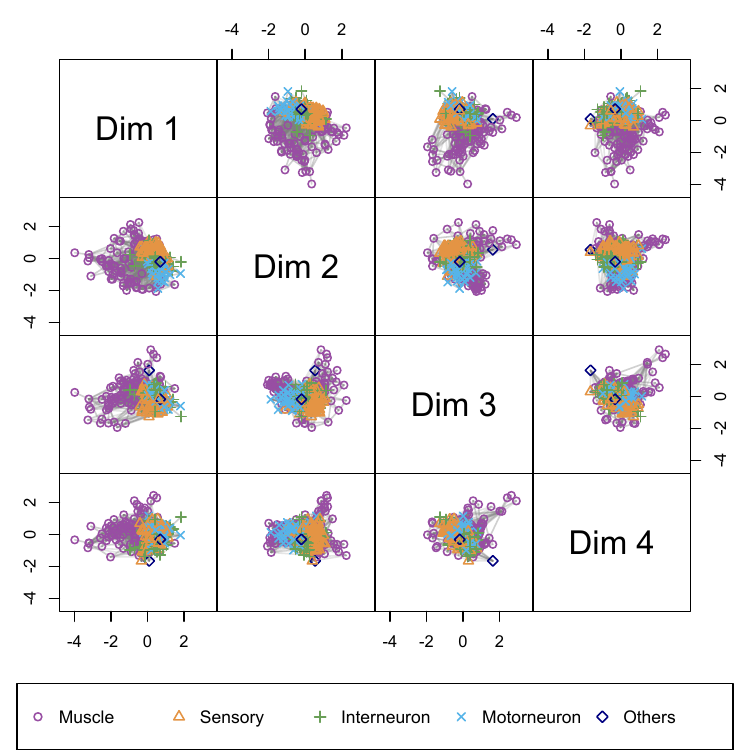}
    \caption{Initial latent positions estimates for the first 4 dimensions under the multidimensional scaling for $p=5$ for the worm nervous system network.}
    \label{fig:vlspm-worm_mds_pos}
\end{figure}

\begin{figure}[htb]
    \centering
    \includegraphics[width=\linewidth]{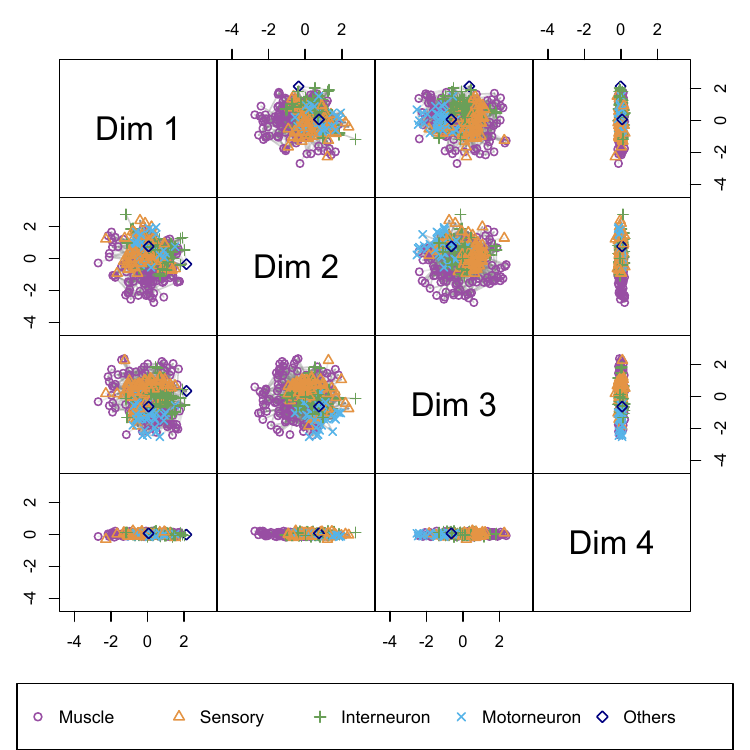}
    \caption{Posterior mean latent positions under the MCMC-LSPM with $p = 4$ for the worm nervous system network.}
    \label{fig:vlspm-worm_lspm_pos}
\end{figure}

Figure \ref{fig:vlspm-worm_mds_pos} shows the initial latent positions for the first 3 dimensions under the multidimensional scaling for $p=5$ for the worm nervous system network while Figure \ref{fig:vlspm-worm_lspm_pos} shows the posterior mean latent positions under the MCMC-LSPM with $p = 3$ for the  worm nervous system network.

\clearpage
\bibliographystyle{apalike}
\bibliography{wileyNJD-APA.bib}{}   

\end{document}